\documentclass[10pt]{article}

\usepackage{amsmath}
\usepackage{array}
\usepackage{appendix}
\usepackage{graphicx}
\usepackage{amsfonts}
\usepackage{amssymb}
\usepackage{mathrsfs}
\usepackage{yfonts}
\usepackage{euscript}
\usepackage{upgreek}
\usepackage{slantsc}
\usepackage{calligra}
\usepackage[T1]{fontenc}
\usepackage{epsf}
\usepackage{latexsym}

\usepackage{tipa}

\usepackage{cancel}

\def\be{\begin{equation}}
\def\ee{\end{equation}}
\def\beq{\begin{equation}}
\def\eeq{\end{equation}}
\def\bea{\begin{eqnarray}}
\def\eea{\end{eqnarray}}

\def\foo{\footnote}

\def\hat{\widehat}

\def\!{\hspace{-1.6667em}}

\def\mD{\mbox{D}}
\def\mE{\mbox{E}}
\def\mF{\mbox{F}}

\def\mJ{\mbox{J}}  

\def\mL{\mbox{L}}
\def\mM{\mbox{M}}
\def\mN{\mbox{N}}

\def\mS{\mbox{S}}
\def\mT{\mbox{T}} 

\def\mV{\mbox{V}}
\def\mW{\mbox{W}}

\def\me{\mbox{e}}

\def\ms{\mbox{s}}
\def\mt{\mbox{t}}

\def\fG{\mbox{\sffamily G}}

\def\fQ{\mbox{\sffamily Q}}

\def\bh{\underline{\underline{\mbox{h}}}  }            
\def\sbh{\underline{\underline{\mbox{\scriptsize h}}}  }     

\def\bM{\mbox{\bf M}}

\def\bM{\mbox{{\bf M}}}
\def\bM{\mbox{{\bf M}}}

\def\bh{\mbox{{\bf h}}}

\def\bq{\mbox{{\bf q}}}

\def\scH{\mbox{\scriptsize ${\cal H}$}}          

\def\scM{\mbox{\scriptsize ${\cal M}$}}          

\def\frM{\mbox{$\mathfrak{M}$}}
\def\sfrM{\mbox{\scriptsize $\mathfrak{M}$}}

\def\sa{\mbox{\scriptsize a}}
\def\sb{\mbox{\scriptsize b}}

\def\sd{\mbox{\scriptsize d}}
\def\se{\mbox{\scriptsize e}}

\def\sg{\mbox{\scriptsize g}} 
 
\def\si{\mbox{\scriptsize i}}

\def\sll{\mbox{\scriptsize l}}  
\def\sm{\mbox{\scriptsize m}}
\def\sn{\mbox{\scriptsize n}} 
\def\so{\mbox{\scriptsize o}}

\def\sr{\mbox{\scriptsize r}}
\def\st{\mbox{\scriptsize t}}

\def\sv{\mbox{\scriptsize v}}
\def\sw{\mbox{\scriptsize w}}

\def\sA{\mbox{\scriptsize A}} 
\def\sB{\mbox{\scriptsize B}}

\def\sD{\mbox{\scriptsize D}}
\def\sE{\mbox{\scriptsize E}}
\def\sF{\mbox{\scriptsize F}}
\def\sG{\mbox{\scriptsize G}}
\def\sH{\mbox{\scriptsize H}}

\def\sJ{\mbox{\scriptsize J}}

\def\sL{\mbox{\scriptsize L}} 
\def\sM{\mbox{\scriptsize M}} 
\def\sN{\mbox{\scriptsize N}} 
\def\sO{\mbox{\scriptsize O}}

\def\sR{\mbox{\scriptsize R}}
\def\sS{\mbox{\scriptsize S}}
\def\sT{\mbox{\scriptsize T}}

\def\sW{\mbox{\scriptsize W}}

\def\sfA{\mbox{\sffamily{\scriptsize A}}}
\def\sfB{\mbox{\sffamily{\scriptsize B}}}

\def\sfZ{\mbox{\sffamily{\scriptsize Z}}}

\def\sbh{\mbox{{\bf \scriptsize h}}}

\def\sbM{\mbox{{\bf \scriptsize M}}}

\def\tb{\mbox{\tiny b}}

\def\te{\mbox{\tiny e}}

\def\tm{\mbox{\tiny m}}
\def\tn{\mbox{\tiny n}}
\def\to{\mbox{\tiny o}}

\def\ttt{\mbox{\tiny t}}   

\def\tw{\mbox{\tiny w}}

\def\tN{\mbox{\tiny N}}

\def\tT{\mbox{\tiny T}}

\def\K{Kucha\v{r} }

\def\pa{\partial}
\def\d{\textrm{d}}

\def\5Star{\mbox{\Large$\star$}}              
\def\sumi3{\sum\mbox{}_{\mbox{}_{\mbox{\scriptsize $i$=1}}}^3}

\def\sumiN{\sum\mbox{}_{\mbox{}_{\mbox{\scriptsize $I$=1}}}^{N}}

\def\sumj3{\sum\mbox{}_{\mbox{}_{\mbox{\scriptsize $j$=1}}}^3}
\def\sumk3{\sum\mbox{}_{\mbox{}_{\mbox{\scriptsize $k$=1}}}^3}

\begin{document}
\begin{titlepage}
\normalfont
\begin{center}

{\bf CLASSICAL MACHIAN RESOLUTION OF THE SPACETIME RECONSTRUCTION PROBLEM}

\mbox{ } 

{\bf Edward Anderson$^1$, Flavio Mercati$^2$}

\mbox{ }

$^1$ {\em DAMTP, Centre for Mathematical Sciences, Wilberforce Road, Cambridge CB3 0WA}, ea212@cam.ac.uk 
 
$^2$ {\em Perimeter Institute, 31 Caroline Street North Waterloo, Ontario Canada N2L 2Y5}, fmercati@perimeterinstitute.ca \normalsize

\end{center}

\begin{abstract}

\noindent Following from a question of Wheeler, why does the Hamiltonian constraint ${\cal H}$ of GR have the particular form it does?.
A first answer, by Hojman, Kucha\v{r} and Teitelboim, is that using embeddability into spacetime as a principle gives the form of ${\cal H}$.
The present paper culminates a second Machian answer -- initially by Barbour, Foster and \'{o} Murchadha -- in which space but not spacetime are assumed.
Thus this answer is additionally a classical-level resolution of the spacetime reconstruction problem.   
In this approach, mere consistency imposed by the Dirac procedure whittles down a general ansatz to one of four alternatives: 
Lorentzian, Galilean, or Carrollian relativity, or constant mean curvature slicing.  
These arise together as the different ways to kill off a 4-factor obstruction term.
It is novel for such an alternative to arise from principles of dynamics considerations 
(in contrast with the historical form of the dichotomy between universal local Galilean or Lorentzian relativity). 
It is furthermore intriguing that it gives constant mean curvature slicing -- 
familiar from York's work on the initial value problem -- as a further option on a similar footing. 
That is related to a number of recent alternative theories/formulations of GR known collectively as `shape dynamics'.
The original work did not treat this with Poisson brackets and a proper systematic Dirac-type analysis; we rectify this in this paper.
It is also the first demonstration of how this approach solves the classical spacetime reconstruction problem via `hypersurface tensor dual nationality' and 
                                                                                                                   what can be interpreted as embedding equations arising.

\end{abstract}

\section{Introduction}

\subsection{GR as geometrodynamics}\label{GR-as-gdyn}

\noindent GR was cast as a dynamical theory in Hamiltonian form by Arnowitt--Deser--Misner (ADM) \cite{ADM} and Dirac \cite{Dirac}.  
Since this formulation is a dynamics of spatial 3-geometry, Wheeler  \cite{Wheeler63, Battelle, MTW} termed it a {\it geometrodynamics}.
It most readily admits a redundant description in terms of the spatial 3-metric, $h_{ij}$. 
This has 6 configurational degrees of freedom, and thus this presentation of GR has 12 phase space degrees of freedom.  
$h_{ij}$ arises as one part of ADM's split of the GR spacetime metric, $g_{\mu\nu}$, as per Fig \ref{7-Actions-2}.b).\footnote{We use lower-case 
Greek and Latin letters for spacetime and spatial indices respectively.
We use ( ) for functions, [ ] for functionals, and ( ; ] for mixed function-functionals. 
That leaves \{ \} without commas for actual brackets. 
We then use bold font to clearly distinguish the Poisson brackets: $\mbox{\bf\{} \mbox{ } \mbox{\bf ,} \mbox{ } \mbox{\bf\}}$.  
The 4-$d$ spacetime is the pair $(\frM, g_{\mu\nu})$. 
$\frM$ is the spacetime manifold at the topological and differentiable structure levels and $g_{\mu\nu}$ a metric that endows it with semi-Riemannian geometrical structure.
$g_{\mu\nu} = g_{\mu\nu}(X^{\rho})$ (for $X^{\rho}$ spacetime coordinates) has determinant $g$, covariant derivative $\nabla_{\mu}$, Riemann tensor $R_{\mu\nu\rho\sigma}^{(4)} = 
R_{\mu\nu\rho\sigma}^{(4)}(x^{\gamma}; g_{\delta\epsilon}]$, Ricci tensor $R^{(4)}_{\mu\nu}$, Ricci scalar $R^{(4)}$ and Einstein tensor $G^{(4)}_{\mu\nu}$.
The 3-$d$ spaces are pairs $(\Sigma, h_{ij})$ all for $\Sigma$ of the same topology.
(Thus such dynamical study restricts $\frM$ to be of the simple form $\Sigma \times I$ for $I$ some kind of interval in $\mathbb{R}$.)  
Moreover, we take this fixed spatial topology to be a compact without boundary one. 
Finally $\Sigma$ additionally comes endowed with suitable differential structure.
$h_{ij} = h_{ij}(x^k)$ (for $x^k$ spatial coordinates) is a spatial metric, with determinant $h$, covariant derivative $D_{i}$, Riemann tensor $R_{abcd} = R_{abcd}(x^e; h_{fg}]$, 
Ricci tensor $R_{ij}$, Ricci scalar $R$ and Einstein tensor $G_{ij}$.  
The block of physical constants occurring as a proportionality in eq (\ref{S-EH}) is set to 1 by choice of units in order to keep the results tidier.}
The other constituent parts of this decomposition are the lapse $\alpha$ and shift $\beta^{i}$. 
These have the status of auxiliary variables that tell one how the 3-spaces are `strutted together'.  


The Einstein--Hilbert action for the conventional spacetime formulation of GR,
\be
\mS_{\sE\sH} = \int_{\sfrM} \d^4x\sqrt{|g|}R^{(4)} \mbox{ } , 
\label{S-EH} 
\ee
can then be rewritten as ADM's action 
\beq
\mS_{\sA\sD\sM} = \int \d\mt \int_{\Sigma}\d^3x \, \mL_{\sA\sD\sM} = \int \d\mt \int_{\Sigma}\d^3x\sqrt{h} \, \alpha \{K_{ab}K^{ab} - K^2   +  R\} \mbox{ } .  
\label{S-ADM}  
\eeq
Here, $\mL_{\sA\sD\sM}$ denotes that formulation's Lagrangian density.
$K_{ab}$ is the extrinsic curvature of the spatial 3-surface.
To understand this conceptually, one first needs to introduce the normal to the spatial 3-surface, $n^{\mu}$.
Then $K_{\mu\nu}$ is the rate of change of $n^{\mu}$ along that spatial 3-surface and thus of its bending relative to the presupposed ambient
4-manifold (the spacetime $\frM$) \cite{Wald},

\end{titlepage}


\beq
K_{\mu\nu} := h_{\mu}\mbox{}^{\rho} \nabla_{\rho} n_{\nu} \mbox{ } .
\label{K-4}
\eeq
Moreover, both the spatial 3-metric and the extrinsic curvature have `dual nationality' as both spacetime and space tensors due to both obeying the {\it hypersurface tensor property}.
Namely, that contractions on each indices with the normal to the hypersurface\footnote{I.e. 
a surface within an ambient space of dimension one greater than that of the surface.} 
vanish: $T_{\mu\nu ... \rho}n^{\mu} = 0$ etc.
It is thus indeed possible to recast the above `spacetime definition' of extrinsic curvature into a spatial hypersurface tensor based ADM-split computational form, 
\beq
K_{ab} =  \{\pa{h}_{ab}/\pa t - \text{\textsterling}_{\underline{\beta}} h_{ab}\}/{2\alpha}  \mbox{ } . 
\label{K-3}
\eeq
Here $t$ is the GR coordinate time, and $\pounds_{\underline{\beta}}$ the Lie derivative with respect to the shift, $\beta^i$.  
The above ADM action obtained by {\sl decomposing} $R^{(4)}$ using a subset of the embedding equations (see Appendix C) for 3-manifolds within 4-manifolds.   


The GR momenta then also turn out to be closely related to the extrinsic curvature, via 
\beq
p^{ab} := \pa \,\mL_{\sA\sD\sM}/\pa \{\pa h_{ab}/\pa t\} = \sqrt{h}\{ K^{ab} - K h^{ab} \} \mbox{ } . 
\label{Gdyn-momenta}
\eeq
I.e. they are a densitized version of $K_{ab}$ with a particular trace term $K := \mbox{tr}(K_{ab})$ subtracted off.  
DeWitt \cite{DeWitt67} then further expressed geometrodynamics in terms of the configuration space geometry.    
The {\it GR configuration space metric} is $\mM^{abcd} := \sqrt{h}\{h^{ac}h^{bd} - h^{ab}h^{cd}\}$. 
This is i) 6 $\times$ 6, ii) indefinite-signature (-- + + + + +) -- most easily seen by using DeWitt's 2-index to 1-index map under which $h_{ij} \rightarrow h^{\sfA}$ and $\mM^{ijkl} 
\rightarrow \mM_{\sfA\sfB}$ -- and iii) ultralocal: no derivative dependence.  
Let us now quantify the redundancy in this 3-metric description.  
4 of the 10 Einstein field equations of GR are constraints\foo{This 
does not require a Hamiltonian analysis to determine, and indeed was figured out long before -- in the 1920's -- by Darmois \cite{Darmois}.}
\beq
\mbox{(GR Hamiltonian constraint) , } \mbox{ } {\cal H}(x^{a}; h_{bc}, p^{de}]   := \mN_{abcd}p^{ab}p^{cd} - \sqrt{h} \, R  = 0  \mbox{ } ,  \label{Ham}
\eeq
\beq
\mbox{(GR momentum constraint) , }    \mbox{ } {\cal M}_i(x^{a}; h_{bc}, p^{de}] := -2D_j p^{j}\mbox{}_i  = 0 \label{Mom}   \mbox{ } ,   
\eeq
The other 6 are evolution equations.  


\noindent Note that the GR Hamiltonian constraint contains the inverse of the GR configuration space metric, 

\noindent $\mN_{abcd} := \{h^{ac}h^{bd} - h^{ab}h^{cd}/2\}/\sqrt{h}$: the {\it DeWitt supermetric}. 
One can also view GR's redundancies from the point of view of it having 10 spacetime metric coordinates and 4 coordinate freedoms.
In the spacetime picture, the conventional underlying symmetries are spacetime 4-diffeomorphisms. 
It is just as simple to view the three coordinate freedoms of 3-space likewise. 
The 3 corresponding first-class (and uncontroversially gauge) constraints then form the GR momentum constraint, ${\cal M}_i$.  
The underlying symmetries are in this case 3-diffeomorphisms, Diff($\Sigma$). 
Taking these into account, one passes from a theory of 3-metrics -- a {\it metrodynamics} -- to a theory of 3-geometries: geometrodynamics.  
3-geometries are taken here to be the `shape information' -- here including a local scale degree of freedom -- 
present in the 3-metric once the unphysical `coordinate grid' information has been removed.
In this paper, we use `pure shape' for notions of shape the explicitly do not include any notions of scale.  
This readily converts to an active view of diffeomorphisms, for which the space in question has no diffeomorphism-invariant notion of points.  

\mbox{ } 

Moreover, the diffeomorphisms of spacetime split up into a foliation by spatial slices are more complicated and less well understood than the diffeomorphisms of spacetime or of space.
The above constraints use up 2 degrees of freedom each, but whether this makes the Hamiltonian constraint ${\cal H}$ a gauge constraint is disputed, as laid out in 
\cite{HT, Kuchar93, Buckets, BF08}.  
For sure, in split form, the symmetries are no longer just the 4-diffeomorphisms, but rather the 3-diffeomorphisms and the hidden refoliation invariance of split formulations.  
For instance, whilst Diff($\frM$) and Diff($\Sigma$) form (infinite-dimensional) Lie algebras, we shall see that the algebraic structure of the GR constraints.
This corresponds to the kind of split in question, is not a Lie algebra but rather an algebroid -- the Dirac algebroid (see Sec \ref{GR-PB}).
Also it remains disputed whether the Hamiltonian constraint is a gauge constraint (which has ties to some parts of the next SSec's Problem of Time). 
For, if it were, it could be taken to imply that the evolution it generates is gauge, rather than physical, so nothing at all would {\sl happen} in a universe thus described. 
Finally, following Wheeler \cite{Battelle} and DeWitt \cite{DeWitt67}, we use Riem($\Sigma$): the space of spatial 3-metrics on a fixed $\Sigma$ (named after Riemann).  
Then use Wheeler's Superspace($\Sigma$) = Riem($\Sigma$)/Diff($\Sigma$): the space of spatial 3-geometries on a fixed $\Sigma$ .

\subsection{The Problem of Time in Quantum Gravity}\label{PoT}

\noindent ${\cal H}$ causes a number of Problem of Time facets (see e.g. \cite{Kuchar92, I93, APOT, APOT2, FileR, BI} for more details). 
The most well-known of these is quantum-level frozenness. 
I.e. a stationary equation $\widehat{\cal H}\Psi = 0$ occurs in a place where one would expect, rather, a time-dependent wave equation (for some notion of time).  
Note more generally that whilst the Problem of Time is commonly perceived as mostly a quantum-level problem, in fact 8 of the 9 Problem of Time facets have classical precursors.

The next Problem of Time facet is purely classical -- the Thin Sandwich Problem \cite{Wheeler63, Sandwich} (see Appendix B for various more general relational versions of this).
This facet involves solely ${\cal M}_i$.   
The third facet is the Constraint Closure Problem: ${\cal H}$ and ${\cal M}_i$ need to close in the algebraic sense under a suitable type of Poisson bracket.  
The fourth facet is the particular difficulty of finding observables/beables.  
`Observables/beables' is meant here in the sense of Dirac are required to commute with ${\cal H}$ and ${\cal M}_i$.
${\cal H}$ and ${\cal M}_i$ are furthermore each related to part of the strutting involved in the Foliation Dependence Problem Facet.
As we shall see below, plays a role in at least the classical Spacetime Reconstruction Problem Facet too.  
The Foliation Dependence Problem concerns whether the physics depends on how spacetime is foliated by spatial hypersurfaces; see the next Subsection for Spacetime Reconstruction.  
Classically, all of the Constraint Closure, Foliation Dependence and Spacetime Reconstruction Problems are resolved by the nature of the Dirac algebroid of the GR constraints 
or extensions of this as considered in the present paper.
The final two Problem of Time facets for GR are the Global Problem of Time and the Multiple Choice Problem, which only occurs upon considering quantum theory.

The above very much motivates studying models in which both ${\cal H}$ and ${\cal M}_i$ are nontrivial.
Moreover, as we shall see, ${\cal M}_i$ arises as an integrability of ${\cal H}$.  
By this all the above consequences of ${\cal M}_i$ can also be considered to originate from ${\cal H}$.

\subsection{Two foundational questions}\label{2FQ}

{\bf Wheeler's Question} The preceding SSec motivates Wheeler's asking the following foundational question, and establishes why it is a particularly important one.  
``{\it If one did not know the Einstein--Hamilton--Jacobi equation, how might one hope to derive it straight off from plausible first principles without ever going through the 
formulation of the Einstein field equations themselves?"} \cite{Battelle}.
This readily translates to asking for first-principles reasons for the form of the GR Hamiltonian constraint, ${\cal H}$.
This is via the usual Principles of Dynamics \cite{Lanczos} inter-relation between a Hamiltonian energy equation and the corresponding Hamilton--Jacobi equation. 

\mbox{ }  

\noindent A first answer to this was given by Hojman, Kucha\v{r} and Teitelboim \cite{HKT76}.  
However, they presuppose spacetime and then get `embedding into spacetime' conditions.  
See Sec \ref{GR-PB} for more about their approach. 

\mbox{ }

\noindent {\bf Classical Spacetime Reconstruction Problem}.  
In this paper, we consider this in the sense of spacetime reconstructed from space.

\mbox{ } 

\noindent Clearly one cannot assume embeddability into spacetime any more if one is looking to answer both of these questions, 
pace Wheeler's {\it``the central starting point in the proposed derivation would necessarily seem to be imbeddability"} (sic) \cite{Battelle}.
He further affirmed this in his later commentary \cite{Wheeler77} of \cite{HKT76}.   
The present paper indeed considers dropping the embeddability assumption, working instead from relational first principles.  
Note the progress from spacetime versus space as primary to spacetime reconstruction from a presentation of space formulated on relational first principles.  
The present paper is the first account for the latter to be submitted for publication. 

\mbox{ }  

\noindent {\bf Quantum Spacetime Reconstruction Problem} \cite{Kuchar92, I93, FileR} (this is beyond the present paper's scope). 
As Wheeler also pointed out \cite{Battelle}, at the quantum level, fluctuations of the dynamical entities are unavoidable. 
In the present case, these are fluctuations of 3-geometry, and these are then too numerous to be embedded within a single spacetime.  
The beautiful geometrical way that classical GR manages to be Refoliation Invariant breaks down at the quantum level.
Wheeler furthermore pointed out \cite{Battelle, Wheeler77} that Heisenberg's uncertainty principle applies.
In GR, this amounts to the quantum operator counterparts $h_{ij}$ and $p^{ij}$ obeying an uncertainty relation. 
But by formula (\ref{Gdyn-momenta}) this means that $h_{ij}$ and $K_{ij}$ are not precisely known.   
Thus the idea of embeddability of a 3-space with metric $h_{ij}$ within a spacetime is itself quantum-mechanically compromised.
Thus (something like) the geometrodynamical picture (considering the set of possible 3-geometries and the dynamics of these) 
would be expected to take over from the spacetime picture at the quantum level.  
It is then not clear what becomes of causality or of locality. 
In particular, microcausality is violated in some such approaches \cite{S-K}.
Additionally, the recovery of semiclassicality aspect of spacetime reconstruction has a long history of causing difficulties in Loop Quantum Gravity \cite{T-FL11-B-D-R}.  
Humankind has also started to investigate the semiclassical and quantum commutator bracket counterpart of the classical Dirac algebroid \cite{Bojo12}.  
The extension of such work to a family of such algebraic structures in parallel with the current paper remains to be tackled.

\mbox{ } 

\noindent {\bf Discrete counterparts of Spacetime Reconstruction Problem}. 
This concerns obtaining classical spacetime or space from more primary classical discrete notions (see e.g \cite{MRS-RiWa}), and also eventually from more primary QM ones.

\subsection{Observing that GR does rest on relational first principles}\label{eph}  

\noindent {\bf Leibnizian Temporal Relationalism} \cite{L}. The sentiment of this can be usefully expressed as `there is no time for the universe as a whole'.
Here no details of what is meant by `universe' are required other than it being a whole system rather than a subsystem. 
(Thus how our conception of universe has changed since Leibniz's day due to Einstein, Hubble and modern observational cosmology does not affect use of this first principle). 

\mbox{ } 

\noindent This principle is to be implemented as follows.

\noindent A) One adopts a geometrical formulation of action principle that happens to be parametrization-irrelevant.\footnote{This is trivially 
mathematically equivalent to reparametrization invariant. 
However, parametrization-irrelevant is conceptually superior through not using a parameter. 
Finally, the geometrical formulation \cite{FileR} is better still through {\sl not even raising} the issue of meaningless parameters!\label{PI-RI-DI} }

\noindent B) One's action is not to contain any extraneous time (e.g. Newton's) or time-like variables (e.g. GR's lapse $\alpha$).

Geometrical actions of this kind for mechanics are due to Jacobi and Synge and are carefully exposited in \cite{Lanczos}.  
For instance, Ordinary Mechanics (in the sense of quadratic-velocity actions or their timeless geometrical reformulations) has the Jacobi action
\beq
\mS_{\sJ} = \sqrt{2} \int\d \mJ = \sqrt{2}\int\d \ms \sqrt{\mW} \mbox{ } . 
\label{Duc}
\eeq 
The configuration space geometrical line element, $\d \ms = \sqrt{\mM_{AB}\d q^A \d q^B} = ||\d \bq||_{\sbM}$.  
$|| \mbox{ } ||_{\sbM}$ here denotes `norm' with respect to the mass matrix, $\bM$.
This line element is most commonly of the form $\sqrt{\sumiN m_I\d \bq^{I\,2}}$.
$\d\widetilde{\ms}$ is the physical line element. 
This differs from the preceding by a conformal factor: the square root of the potential quantity, $\sqrt{\mW}$, for $\mW := \mE - \mV$, V the potential energy and E the total energy.
In the case of GR, one has in place of this the Barbour--Foster--\'{o} Murchadha--A (BFO--A) action \cite{RWR, San, FileR}.   
[This is similar to the more familiar Baierlein--Sharp--Wheeler (BSW) \cite{BSW} action.    
See (\ref{S-rel}) for the form of this and Sec \ref{Demon} for the differences between it and its mechanical counterpart.]
Such actions have two consequences.   

\noindent 1) They are known (via \cite{BB82} and footnote \ref{PI-RI-DI}) to encode primary constraints.  
This yields an energy constraint for Mechanics and the famous GR Hamiltonian constraint ${\cal H}$ for the GR case. 
These are all quadratic and not linear in the momenta.

\noindent 2) A particular emergent time ({\it emergent Jacobi time}, $t^{\se\sm(\sJ)}$) happens to simplify the momentum--velocity relations and Euler--Lagrange equations of motion.
Adopting this for this reason leads, in the case of Mechanics, to an entity numerically equal to, but conceptually different from, the Newtonian time. 
The same procedure produces proper time for GR in general and cosmic time for simple GR cosmology models.  
Moreover this emergence implements Mach's Time Principle `time is to be abstracted from change' \cite{M, ARel2}, as seen most clearly for the Mechanics example 
\beq
t^{\se\sm(\sJ)} = \int{\d \ms}\left/\sqrt{2\mW} \right.  \mbox{ } .
\label{t-J}
\eeq
\mbox{ } \mbox{ } N.B. that this is manifestly of the form d(change).  
A forteriori, it is of the form d(all change) in principle \cite{B-Essay}, but d(STLRC) in practice \cite{ARel2}, STLRC standing for `sufficient totality of locally relevant change'. 
Times of this type gain inspiration from the astronomers' ephemeris time concept \cite{Clemence}. 
$t^{\se\sm(\sJ)}$ is a `Generalized Local Ephemeris Time' (GLET).  
This embodies the `Machian Democracy Principle' of giving every change the opportunity to contribute whilst having the practical bonanza of then rejecting those changes that 
do not contribute to that desired accuracy.  
Rovelli \cite{R-Essay, Rovellibook} often considers {\sl any} change. 
The GLET is a {\sl distinct} distinguished entity that actually corresponds to the generalization of the sequence of procedures used for accurate timekeeping on Earth from Ptolemy 
until the 1970's.
Ephemeris time itself was introduced in the mid-20th century, with atomic clocks subsequently calibrated against accurate Celestial Mechanics observations.
More generally, see Appendix A for a summary of the Temporal Relationalism incorporated Principles of Dynamics (TRiPoD), 
and Appendix B for how to interpret extrinsic curvature in this context.

\mbox{ } 

\noindent {\bf Configurational Relationalism} concerns configuration spaces $\fQ$ that have acting upon them groups of physically-irrelevant transformations $\fG$.  
This covers both 1) the internal symmetries of conventional gauge theory: Internal Relationalism. 
2) Transformations of space itself (translations and rotations of absolute space for relational mechanics, or spatial 3-diffeomorphisms for GR: Spatial Relationalism.  
The {\sl indirect implementation} of Configurational Relationalism involves using not $\fG$-redundant objects $O$ but rather 
\beq
\mbox{\Large S}_g \circ \stackrel{\rightarrow}{G}_g O \mbox{ } . 
\eeq 
Here $\stackrel{\rightarrow}{G}_g$ is the action of group $\fG$, $\circ$ denotes composition of maps, 
and $\mbox{\Large S}_g$ is an `all' move that renders the whole construct $\fG$-irrelevant. 
Examples of `all' moves are 
i)   group averaging or integrating over a group, 
ii)  taking infs or sups over a group, 
iii) carrying out an extremization over a group.  
A particular case of the last of these -- `Best Matching' -- is used to form and at least in principle process the actions used in Barbour's relational program.  
This works by producing a constraint from variation with respect to the auxiliary $g^{\sfZ}$, which uses up 2 degrees of freedom per $g^{\sfZ}$ degree of freedom 
(I.e. the $g^{\sfZ}$ themselves and one $\fQ$ degree of freedom per $g^{\sfZ}$ degree of freedom, in the manner familiar of gauge constraints \cite{HT}.) 
Thus the extremization sends one to the quotient space $\fQ/\fG$ as desired.  
This involves the solution of the linear constraints Lin$_{\sfZ}$ for their own auxiliary variables $g^{\sfZ}$. 
This is in fact need prior to obtaining an explicit generalization of (\ref{t-J}).
This is because one now picks up an extremization over group elements $g^{\sfZ} \in \fG$ outside of its action, in which case it is termed the Jacobi--Barbour--Bertotti emergent time, 
$t^{\se\sm(\sJ\sB\sB)}$ \cite{BB82}.  
Then for 1- and 2-$d$ relational mechanics, Best Matching is unconditionally solvable so $t^{\se\sm(\sJ\sB\sB)}$ can be given an explicit expression \cite{FileR}. 
For GR, however, this extremization is none other than the notorious and largely-unresolved Thin Sandwich Problem \cite{Sandwich}.    
Here, $\fQ$ = Riem($\Sigma$), $\fG$ = Diff($\Sigma$) and Riem($\Sigma$)/Diff($\Sigma$) = Superspace($\Sigma$).


Note that one needs to use d$g^{\sfZ}$ in place of $g^{\sfZ}$ so that the Configurational Relationalism does not spoil the parametrization irrelevance that implements 
Temporal Relationalism.  
This is successful \cite{ABFO, FEPI, FileR}, and involves d$F^i$ to be used in place of the more common shift $\beta^i$.\foo{$\d F^i$ 
is the cyclic differential of the {\it frame}, $F^i$, which is the manifestly parametrization irrelevant counterpart of the cyclic velocity of the frame $\dot{F}^i$.  
In turn, this is the manifestly reparametrization invariant replacement for the ADM Lagrange multiplier coordinate shift $\beta^i$. 
See Sec \ref{Demon} and Appendix A for more details about this progression.}

That is indeed the sole difference between the BSW and BFO--A actions.  
For now, the geometrical form of the latter is (using a bar to denote densitization) 
\beq
\mS^{\sG\sR}_{\sr\se\sll\sa\st\si\so\sn\sa\sll} = \int\d\lambda\int_{\Sigma}\d^{3}x\sqrt{\bar{R}} \, \d \ms^{\sG\sR}_{\sr\se\sll\sa\st\si\so\sn\sa\sll} 
\mbox{ } \mbox{ }  \mbox{ } \mbox{for } \mbox{ } 
\d \ms^{\sG\sR}_{\sr\se\sll\sa\st\si\so\sn\sa\sll} := ||\d_{\underline{F}} {\mbox{\boldmath$h$}}||_{\sbM} 
                                                    = ||\d {\mbox{\boldmath$h$}} - \pounds_{\d\underline{F}} \mbox{\boldmath$h$}||_{\sbM} 
\label{S-rel} \mbox{ } .  
\eeq
The usefully shortened notation above has the following meaning.  

\noindent 1) $||\mbox{ } ||_{\sbM}$ is now not a true norm because $\bM$ is the undensitized GR configuration space metric and this is indefinite.

\noindent 2) $\d_{F}$ is a {\it best-matched} differential, which is the most important part of satisfying Configurational Relationalism.

\noindent Then indeed manifest parametrization irrelevance guarantees \cite{Dirac} GR's Hamiltonian constraint (\ref{Ham}) as a primary constraint.
Also, indeed variation with respect to the auxiliary Diff($\Sigma$)-variables $F^i$ gives the GR momentum constraint (\ref{Mom}).  
These constraints close, trivially, due to the Bianchi identity (see e.g. \cite{Wald}).  
Finally GR's classical Machian emergent Jacobi--Barbour--Bertotti time \cite{B94I, SemiclI, ARel2} is
\beq
t_{\sG\sR}^{\se\sm(\sJ\sB\sB)}(x^i) =  \int ||  \d_{F_{\tb\tm}}  {\bf h} ||_{\bf M} \left/ \sqrt{\bar{R}} \right. 
\eeq
for $F_{\sb\sm}$ the best-matching extremum of $S^{\sr\se\sll\sa\st\si\so\sn\sa\sll}_{\sG\sR}$ over the $\mbox{Diff}({\bf \Sigma})$.  
See \cite{GrybTh-Pooley, B11, ARel, FileR, Rovellibook, Kieferbook} for further reviews of various forms of Relationalism and comparisons between them.

\subsection{Relational first principles lead to answers to both questions}

\noindent The basic point is deriving GR as one of a number of possibilities that are relational, explaining all else that comes out in the process (Sec \ref{RWR}).  
Suppose one adopts the relational first principles {\sl without} assumption of additional features derived in ADM's approach.  
Then this leads to the {\sl recovery} of the BFO--A action of GR as one of very few consistent choices within a large class of such actions. 
This `relativity without relativity' (RWR)\footnote{`X without X' 
is a frequently-used phrase of Wheeler's concerning obtaining entity X from deeper first principles.} 
program was built up in \cite{RWR, AB} and compared with Hojman--Kucha\v{r}--Teitelboim's approach in \cite{Phan}. 
This involves not assuming the GR form of the kinetic metric or of the potential. 
As we lay out in Sec \ref{RWR}, One then has a family of trial actions with kinetic arc element ansatz built out of 
the general ultralocal supermetric $\mM_{\st\sr\si\sa\sll}\mbox{}^{ijkl} := \sqrt{h}\{h^{ik}h^{jl} - w h^{ij}h^{kl}\}/y$ 
and potential density ansatz       $\bar{\mW}^{\st\sr\si\sa\sll} = \sqrt{h}\{a + b R\}$ for $a$, $b$ constant,
\beq
\mS_{\st\sr\si\sa\sll} = \int\int_{\Sigma}
\d^3x \sqrt{ \bar{\mW}^{\st\sr\si\sa\sll}}\d \ms^{\st\sr\si\sa\sll}  \mbox{ } .  
\eeq
Then relational postulates alongside a few mathematical simplicities already give this 
since the Dirac procedure \cite{Dirac} prevents most other choices of potential term $\d \ms^{\st\sr\si\sa\sll}$ from working \cite{RWR, San, Than, Lan, Phan}.  
It also fixes $w$ to take the DeWitt value 1 of GR as one of very few consistent possibilities.
Then the constraint propagation gives a term with four factors \cite{San, Lan, Than, Phan}. 
These correspond to the three qualitatively different relativities -- Galilean, Lorentzian and Carrollian (infinite, finite and zero speed of propagation). 
Moreover, they arise alongside a surprise fourth factor: constant mean curvature (CMC) slicing.
This might be viewed as part of a refoliable theory (GR) or as a unique foliation that is a new manifestation of absolute simultaneity.\footnote{See 
\cite{Gomes13} for a recent update on the outcome of attempting to further include higher-curvature terms in the potential. 
Also $\mM, \d s^2$ are, strictly, overlined quantities, and $\d \widetilde{s}^2$ is a double-overlined quantity (weight-2 density).}

\subsection{This paper's methodology in completing the above relational answer}

A serious limitation in the original RWR papers was identified in \cite{FileR}, but is only addressed for the first time in the present paper. 
Constraints do not just arise from the form of the action or directly from variation; the constraints that arise in these ways can give rise to further constraints. 
[This has already been mentioned in Sec \ref{PoT} as part of the Constraint Closure Problem.]
The original RWR papers then aim to use that enough constraints can render a theory trivial -- by using up all of its degrees of freedom -- or inconsistent -- 
by imposing yet further restrictions.
However, the simple method employed in \cite{RWR, AB} to investigate constraint propagation did not properly take into account the possibility of second-class constraints.
First-class constraints use up 2 degrees of freedom each, whilst second-class ones use up only one each.
Both are common in Theoretical Physics, so establishing uniqueness by the hypothetical alternative theories picking up enough constraints to be trivial or inconsistent 
is only really of interest if the procedure used to do so does not preclude second-class constraints. 
That envisaged, the procedure is only correct if it adheres to the {\sl worst}-possible case scenario that each constraint use up only one degree of freedom 
until it is finally demonstrated that the constraint has succeeded in remaining first-class by then end of the Dirac procedure.
Understanding why such a demonstration is not straightforward requires the Poisson Brackets formulation.  
First-class constraints weakly close under Poisson brackets (`weakly' means up to functionals of the constraints), whereas second-class ones do not. 
There is however a problem with making this diagnostic at any particular step in Dirac's systematic method of study within which constraints can give rise to further constraints. 
I.e. that a constraint which closes with all constraints that are known at that stage may nevertheless cease to close with a {\sl subsequently} found constraint.
Thus one has to regard constraints as potentially only using up one degree of freedom each until {\sl all} the constraints have been found.
The present paper derives and interprets these `constraint propagation' results in the more rigorous manner afforded by the Poisson brackets formulation. 
For this, general standard procedures along the lines of Dirac's \cite{Dirac} are available, and from these further insights ensue.

Thus we end consideration of the case of classical RWR for geometrodynamics and token matter fields as are needed to have a bona fide recovery of local theories of relativity.
We show that a Dirac-type procedure suffices to obtain the three local theories of relativity corresponding to zero, finite and infinite signal propagation speed.
These result from the strong vanishing of each of a product of three cofactors (`strong' here means by fixing numerical coefficients).  
They arise alongside a CMC-sliced option.
We additionally consider what happens if metrodynamics alone is assumed. 
This drives one back toward GR, a CMC slicing, or toward distinct 5 degrees of freedom per space point versions of the zero and infinite propagation speed theories. 
A theory with 3 degrees of freedom that lies between a metrodynamics and a geometrodynamics arises as a final and unexpected possibility from this working.  


We also cast RWR as a type of Spacetime Reconstruction: a bottom-up approach as opposed to a top-down approach with spacetime assumed.
(ADM and to a lesser extent Hojman--Kucha\v{r}--Teitelboim, who assumed the existence of spacetime among their conditions for determining the form of the Hamiltonian constraint.)   
We tie this down with careful two-way interpretation of embedding/projection equations in Appendix C.  
The Spacetime Reconstruction Problem is itself a Problem of Time facet. 
By providing this here based on a generalized ansatz family of theories' expansion on structures along the lines of the Dirac algebroid formed by the GR constraint,   
we finish showing that an extended family of such algebroids contains all the magic necessary to overcome three Problem of Time facets at the classical level.  
Namely the Constraint Closure Problem, the Foliation Dependence Problem and the Spacetime Reconstruction Problem are jointly resolved at the classical level 
by the form of the Dirac algebroid.  


Finally, we further motivate this paper as a precursor to investigating the semiclassical version of Teitelboim's classical resolution \cite{T73} of Refoliation Invariance.
(This is easier than fully quantum ones and in line with the Problem of Time program in \cite{ARel2, ACos2AHall, FileR} as probably best summarized for now in \cite{BI}).  


\section{Demonstration that GR is a relational theory}\label{Demon}

\subsection{Various presentations of geometrodynamics}
%
{            \begin{figure}[ht]
\centering
\includegraphics[width=0.97\textwidth]{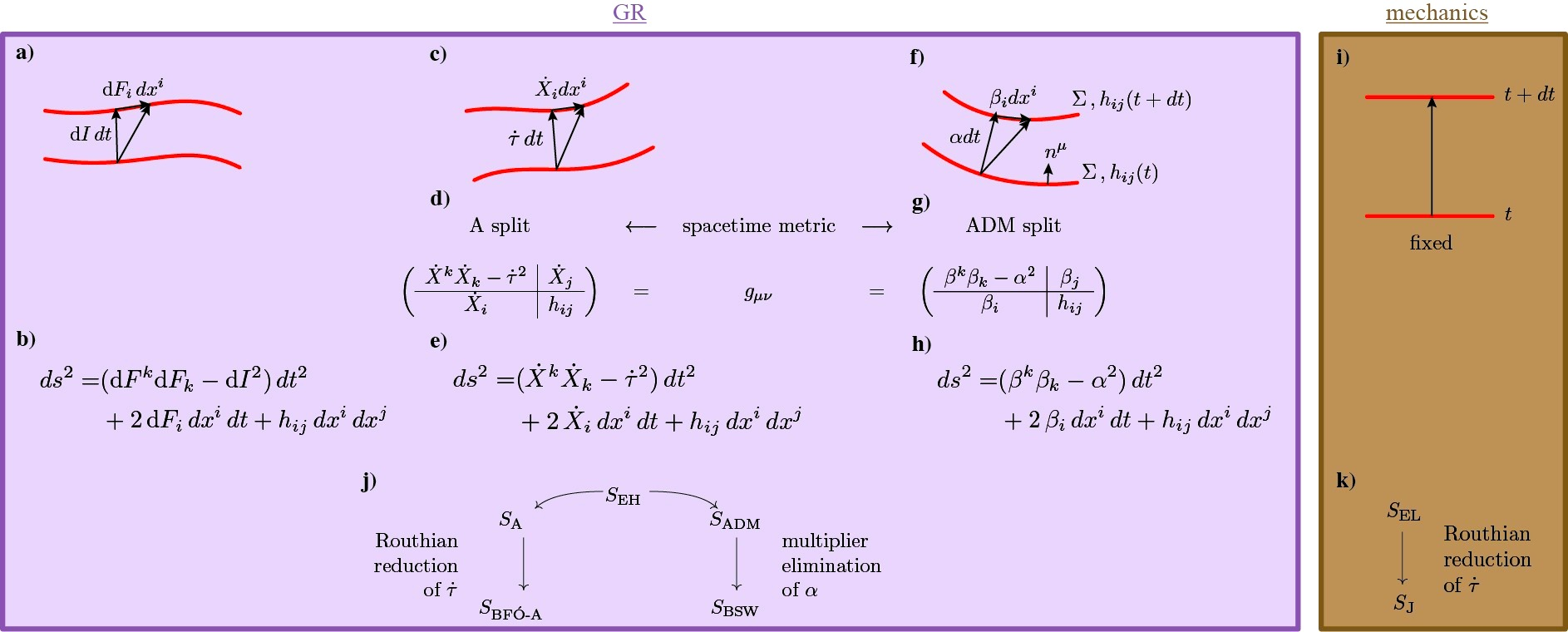} 
\caption[Text der im Bilderverzeichnis auftaucht]{        \footnotesize{In this paper, we start with the entities in the first column a, b). 
Previous conceptualizations are in the second and third (ADM) columns. 
The mechanical counterpart is in the fourth column.  
g, h) is the ADM split, with geometrical significance f). 
d, e) is the A split, with geometrodynamical significance c).  
i) is the mechanics counterpart of these geometrical significances.  
j, k) gives the relations between the seven actions that feature in this paper.} }
\label{7-Actions-2}\end{figure}            }

\noindent Let $\Sigma_1$ and $\Sigma_2$ be neighbouring spatial hypersurfaces within a foliation of spacetime (so they do not intersect).
These are in effect labelled by proper time values $\tau_1 := \tau$ and $\tau_2 := \tau_1 + \d \tau = \tau + \d \tau$, 
and are coordinatized by                           $ x^i_1 := x^i$  and $x^i_2  :=  x^i_1 + X^i     = x^i  + X^i$
Thus $X^a$ is the `difference in coordinate grid' between the two neighbouring spatial hypersurfaces.  
Then only very slightly repackaging the concept on p 507 of \cite{MTW}, one obtains the `A-split'
\beq
ds^2 = \mbox{(proper distance along $\Sigma_2$)}^2 - \mbox{(proper time from $\Sigma_1$ to $\Sigma_2$)}^2 = h_{ij}\{\d x^i + \d X^i\}\{\d x^j + \d X^j\} - \d\tau^2  
\eeq
as depicted in Fig 1.a).  
The first term in this is obtained via $||\d x_2||_{\sbh}\mbox{}^2$.  
If one then parametrizes this A-split with respect to a label time $\lambda = t$ (a GR {\it coordinate time}), one obtains the arc element and matrix in column 2 of Fig 1 \cite{FEPI}.
Suppose one furthermore {\sl defines} {\it lapse} $\alpha$ to be $\d \tau/\d t$ [i.e. d(proper time)/d(coordinate time)], 
                             and {\it shift} $\beta^i$ to be $\d X/\d t$   [i.e. d(difference in coordinate grid)/d(coordinate time)]. 
Then one arrives at the more familiar Arnowitt--Deser--Misner (ADM) \cite{ADM} split of column 3 of Fig 1.  
This is what is usually used, and yet \cite{FEPI, FileR} demonstrates that introducing these quantities is both unnecessary and furthermore not manifestly relational.
That will be explained in due course, but our account considers first a number of features of the familiar formulation, so as then to transcribe them into the relational formulation.   

The conventional and historical spacetime presentation of GR's Einstein--Hilbert action (\ref{S-EH}) is well-known to decompose under the ADM split to give the ADM action. 
The {\sl manifestly}-Lagrangian form for this is 
$$
\mS_{\sA\sD\sM-\sL} = \int \d\mt \int_{\Sigma}
\d^3x\alpha \sqrt{h}\big\{\{h^{ac}h^{bd} - h^{ab}h^{cd}\}\{\pa h_{ab}/\pa t - \pounds_{\underline{\beta}}h_{ab}\}\{\pa h_{cd}/\pa t - \pounds_{\underline{\beta}}h_{cd}\}/4\alpha^2 + R\big\} 
$$
\beq
:=  \int \d\mt \int_{\Sigma}\d^3x\alpha\big\{\bar{\mT}_{\sA\sD\sM-\sL}/4\alpha^2 + \bar{R}\big\}
\mbox{ } .
\label{S-ADM-L}  
\eeq
The lapse $\alpha$ and the shift $\beta^i$ here play the roles of Lagrange multiplier coordinates in the variational procedure entailed by the standard Principles of Dynamics 
treatment of this action.  
Moreover, one can eliminate $\alpha$ from its own multiplier equation readily, on account of this being a very simple algebraic equation: $\alpha^2 = \mT_{\sA\sD\sM-\sL}/4R$.

Performing this elimination gives the Baierlein--Sharp--Wheeler (BSW) action, 
\beq
\mS_{\sB\sS\sW} = \int \d\lambda \int \d^3x \sqrt{ \bar{R} } \sqrt{\bar{\mT}_{\textrm{BSW}}}  \mbox{ } , \label{S-BSW} 
\eeq
where the BSW kinetic energy density is now 
\be
\bar{\mT}_{\textrm{BSW}} = \sqrt{h}\{h^{ac}h^{bd} - h^{ab}h^{cd}\} \{ \pa{h}_{ab}/\pa\lambda - \text{\textsterling}_{underline{\beta}} h_{ab} \}
                                                     \{ \pa{h}_{cd}/\pa\lambda - \text{\textsterling}_{\underline{\beta}} h_{cd} \} \mbox{ } . 
\ee
Here $\lambda$ is a meaningless label (though for GR $\lambda$ and $t$ coincide due to GR being an {\it already-parametrized} theory).
Thus the distinction between this kinetic term and the preceding one is, for GR, entirely a matter of conceptualization rather than of mathematical difference.  
This expression can furthermore be shortened using 
a) the configuration space metric geometry notation for GR of Sec \ref{GR-as-gdyn}.
b) The `norm' notation of Sec \ref{eph}.
c) In the present context of presupposing that one's spatial hypersurfaces are embedded within a manifold of dimension 1 higher -- usually GR spacetime, 
the combination $\pa/\pa \lambda - \text{\textsterling}_{\beta}$  may be conceived of as a {\it hypersurface derivative} \cite{Kuchar76}:  
\beq
\delta_{\stackrel{\rightarrow}{\beta}} := \pa/\pa t - \pounds_{\underline{\beta}} = \pa/\pa\lambda - \pounds_{\underline{\beta}} \mbox{ } .  
\eeq
Here, $\stackrel{\rightarrow}{\beta} = [1, \underline{\beta}]$: a 4-vector in the presupposed spacetime.
Then in terms of this,
\be
\bar{\mT}_{\textrm{BSW}} = \left|\left|\delta_{\stackrel{\rightarrow}{\beta}}\mbox{\boldmath$h$}\right|\right|_{\sbM}^{\mbox{ }\mbox{ } 2} \mbox{ } . 
\ee
\mbox{ } \mbox{ } Note that Temporal Relationalism is {\sl not} satisfied for the BSW action.
It does succeed in being formulated free from an extraneous time-like notion such as the GR lapse (by which the ADM action is {\sl even less} temporally relational). 
And it would have cancellation between its dots and the $\d \lambda$, except that the shift multiplier coordinate vector $\beta^i$ breaks manifest reparametrization-invariance.
This can be avoided by using a formulation in which a shift is not defined in the first place.
One can attain this by working with the difference in coordinate grid frame $F^i$ alongside a lapse $\alpha$.
However, this would be i) a conceptually-mixed formulation and ii) still not in line with further insights from Mechanics as regards the implementation of Temporal Relationalism.

Before proceeding, let us clarify ii).  
In Mechanics, Temporal Relationalism can be acquired by moving from the Euler--Lagrange action 
\be
\mS_{\sE\sL} = \int \d t^{\sN\se\sw\st\so\sn} \{\mT_{\sE\sL} - \mV(\mbox{\boldmath$q$}) \} \mbox{ } \mbox{ } \mbox{for } \mbox{ } \mbox{ } 
\mT_{\sE\sL} = \sumiN m_I\{\pa q_I/\pa t^{\sN\se\sw\st\so\sn}\}^{2}/2 \mbox{ } ,   
\ee
to the Jacobi action\footnote{In this mechanics case $\lambda$ and $\mt$ are {\sl not} the same: mechanics is {\sl not} already-parametrized. 
The role of $t$ is played by $t^{\tN\te\tw\ttt\to\tn}$, whereas the Jacobi formulation makes no primary mention of any such quantity.}
\be
\mS_{\sJ} =  2\int\d \lambda\sqrt{ \mT_{\sJ}\{\mE - \mV(\mbox{\boldmath$q$})\} } \mbox{ } \mbox{ } \mbox{ for } \mbox{ } \mbox{ } 
\mT_{\sJ} = \sumiN m_I \{\pa{q}_I/\pa\lambda\}^{2}/2 \mbox{ } .
\label{flick}
\ee
This is by the process known as {\it passage to the Routhian} alias {\it Routhian reduction} \cite{Lanczos}, which involves elimination of a {\it cyclic velocity}. 
Thus, while it bears close parallels as an elimination with the above BSW Lagrange multiplier elimination, the analogy is not exact 
so long as one formulates one's action in terms of a lapse Lagrange multiplier coordinate.
However, if one uses the A-split of Fig 1.b), the coordinate-time derivative of the proper time $\tau$ takes the place of the lapse in the action, 
\beq
\mS_{\sA}= \int \d\mt \int_{\Sigma}\d^3x  \,  \frac{\pa\tau}{\pa t}  
\left\{ 
\mbox{\Large||}\delta_{\frac{\pa X}{\pa t}} \bh\mbox{\Large||}_{\sbM}
\left/ 
4
\left\{
\frac{\pa\tau}{\pa t}
\right\}
\right.^{2}
+ \bar{R}
\right\} \mbox{ } .   
\label{S-A}  
\eeq
Thus, using this, the analogy between GR and mechanics analogy becomes exact.
Now a `Routhian reduction to remove $\pa\tau/\pa t$' replaces BSW's `multiplier elimination to remove $\alpha$'.

Note that this is a nontrivial change of formalism since it changes the set of variables that are to be identified as those that one is to carry out variations with respect to.
Thus this is not {\sl just} applying\footnote{In fact, this is a case of {\sl un}applying: 
by never making and instead expressing one's action in terms of more primitive variables with respect to which one is then to vary.} 
the definition of lapse (and likewise for the shift). 
The outcome of doing this is that one needs to consider a more subtle variational procedure due to the interplay between the unphysicality of auxiliary variables and these being cast 
as the cyclic velocities of the variables to be varied. 
The ensuing `free end spatial hypersurface value' variation \cite{FEPI} then turns out to return the usual (ADM) equations of GR.  
Thus casting GR in temporally relational form does not alter the form of its equations.\footnote{Thus, as a useful check, 
the relational approach maintains that ADM had the right equations implied by relationalism. 
Moreover, this is from a non-relational argument which can nevertheless be recast as a correctly-relational argument as outlined above.}

Also note that setting up action (\ref{flick}) requires a slightly different formulation of extrinsic curvature. 
I.e. not (\ref{K-3}), which recast in terms of the hypersurface derivative is  $K_{ab} = \delta_{\stackrel{\rightarrow}{\beta}}h_{ab}\big/2\alpha$ but rather 
(using $\dot{\mbox{ }} = \pa/\pa\lambda$)  
\beq
K_{ab} = \left. \delta_{\stackrel{\rightarrow}{\dot{X}}}h_{ab} \right/ 2\{\pa{\tau}/\pa t\} \mbox{ } .   
\label{lis}
\eeq
\noindent Then from (\ref{S-A}) we arrive at
\be
\mS_{\sB\sF\sO-\sA} = \int \d\lambda \int \mbox{d}^3x  \sqrt{ \bar{R} } \sqrt{\bar{\mT}_{\sB\sF\sO-\sA}} . \label{S-BFO-A}
\ee
Here the kinetic energy density is now considered to be of the form
\beq
\bar{\mT}_{\sB\sF\sO-\sA} = \left|\left|\delta_{\stackrel{\rightarrow}{\dot{X}}}\mbox{\boldmath$h$}\right|\right|_{\sbM}^{\mbox{ }\mbox{ } 2} \mbox{ } .  
\eeq
Now indeed Configurational and Temporal Relationalism are implemented without interfereing with each other.

\mbox{ } 

There are three further upgrades in conceptualization to make. 

\noindent 1) Postulate actions of this kind from relational first principles rather than as the outcome of presupposing spacetime, 
A-decomposing it and then applying Routhian reduction.
One marks this as follows.

\noindent i) Implement Configurational Relationalism using frame variable auxiliaries, an example of which is a vector $F^i$ (in the role played so far by $X^i$).

\noindent ii) Use a distinct name and symbol for the emergent quantity that supplants the presupposed $\tau$. 
We use the symbol `$I$' for `instant' since this emergent entity is, most primarily, a {\it labeller of instants}. 
This turns out to play the role of proper time, so that `instants are labelled by proper time', which is very satisfying.

\noindent 2) One can avoid mentioning meaningless labels at all (the first novel component of the TRiPoD).  
This is because one can use $\d F^i$ rather than $\dot{F}$, actions built out of Jacobi geometrical arc elements (as per the Introduction) in place of Lagrangians.  
Furthermore once one has such an emergent entity, $\d I$ in place of $\dot{I}$.
We can think of this as manifest parametrization irrelevance rather than manifest reparametrization invariance.   
It is even better simply to consider a geometrical action conceived of during which parametrization issues never even occurred. 
One would then name the approach after this procedure.  
This is as opposed to after what should have never been considered at the primary level in physical theory in the first place during the conceptual development of this approach. 
I.e. meaningless label times. 

\noindent 3) One is no longer to think in terms of spatial hypersurfaces presupposedly embedded into a manifold of dimension one higher such as GR spacetime. 
Instead, it is simplest and most natural to think of the $\d_{\underline{F}} := \d - \pounds_{\d \underline{F}}$ combination that is left in the role originally played by 
$\pa/\pa\lambda - \pounds_{\underline{\beta}}$ as a {\sl best matching derivative}. 
N.B. the latter is a concept about 3-spaces themselves.\footnote{There is no problem in further envisaging 
this for $n$-spaces of any other dimension, but what is {\sl not} involved here is a combination of space of different dimension.  
This is quite unlike in the study of $n$-spaces that are presupposed to be hypersurfaces embedded within a manifold with one dimension more and one has a geometrical structure of 
already-known type.}
%
That these two derivatives are mathematically the same rests upon the spacetime--space duality of hypersurface objects.   
(This is regardless of whether this hypersurface-ness is assumed or emergent.) 
This is most easily seen in the intermediate BFO--A formulation.  
Here, $\delta_{\stackrel{\longrightarrow}{\pa{\underline{F}}/\pa\lambda}} = \pa/\pa\lambda - \pounds_{\pa \underline{F}/\pa\lambda} := \bullet_{F}$, and the ensuing action is
\beq
\mS_{\sB\sF\sO-\sA}  = \int \d\lambda \int \mbox{d}^3x \sqrt{ \bar{R}} \sqrt{\bar{\mT}_{\sB\sF\sO-\sA}}  \mbox{ } \mbox{ } , \mbox{ } \mbox{ } \mbox{ } 
\bar{\mT}_{\sB\sF\sO-\sA}  = \left|\left|\bullet_{\underline{F}}\mbox{\boldmath$h$}\right|\right|_{\sbM}^{\mbox{ }\mbox{ } 2} \mbox{ } .  
\label{Action-BFO-A}
\eeq
\noindent However, if we apply all of 1), 2), 3), we arrive, finally by this epistemic account, but first of all from what we consider to be the full application of the relational 
first principles, at action (\ref{S-rel}).

\subsection{Relationalism requires a $\mbox{\boldmath$differential$-$almost$}$-Hamiltonian formulation}\label{DAH}

In reformulating the above in Poisson brackets for the above, we start by making two further TRiPoD developments.

\noindent 1) Instead of involving a                              {\it total    Hamiltonian} built by appending constraints with Lagrange Multiplier coordinates, 
manifest Temporal Relationalism requires a                       {\it total dA-Hamiltonian} built by appending constraints with cyclic differentials \cite{FileR}.
\cite{FEPI} considered the intermediate notion of   total                    A-Hamiltonian built by appending constraints with cyclic velocities).\footnote{Please note 
the hyphenation -- the concept in use is a total `d{\it A}-Hamiltonian' rather than any kind of `total differential'.
Here and elsewhere in this paper `{\it A}-' stands for almost- and `d{\it A}' stands for differential-almost-.}
%
It is an almost-Hamiltonian because, whilst it depends purely on momenta for physical variables, it departs from a Hamiltonian via its dependence on auxiliary variables' velocities. 
I.e. it is the same concept as regards its physical content but a more general concept as regards its auxiliary content. 
[This is similar to the spirit in which Dirac introduced multiple notions of Hamiltonians 
up to what types of constraint terms were appended unto them with what type of Lagrange multipliers.]
In the present GR case, this is $\d A_{\sT\so\st\sa\sll} = \d I    \, {\cal H} + \d F^i {\cal M}_i$ in place of, successively, 
                                   $A_{\sT\so\st\sa\sll} = \dot{I} \, {\cal H} + \dot{F}^i {\cal M}_i$ and 
								   $H_{\sT\so\st\sa\sll} = \alpha  \, {\cal H} + \beta^i   {\cal M}_i$.  
Note that these entities differ in no way as regards the {\sl physical} information encoded. 
In particular the objects containing the physical degrees of freedom are encoded as (configuration, conjugate momentum) pairs on the usual grounds.  

\noindent 2) We comment that the Dirac procedure allows for 4 types of equation to arise upon taking Poisson brackets of constraints. 
I.e. inconsistencies, identities, constraints and equations specifying Lagrange multiplier coordinates (`specifier equations'). 
These four cases are well known \cite{HT} to be able to arise {\sl in any combination at any level} of the Dirac procedure.
The Temporal Relationalism-implementing d{\it A}-Dirac procedure counterpart then allows likewise for 4 types of equation to arise in any combination at any level.
This has the induced obvious difference that `specifier equations' now specify cyclic differential auxiliaries rather than Lagrange multiplier ones.

\subsection{The relational system of equations for GR}

The result of varying the manifestly-relational formulation of the GR-as-geometrodynamics action (\ref{S-rel}) is as follows.
We write this using 
\be
\d I := \d \ms\left/2\sqrt{ \bar{R} }\right.  
\ee 
for this scheme's emergent differential of the instant (DOTI), $\d I$ that simplifies the scheme's equations of motion.

In the true relational formulation this supplants the Lagrangian and therefore enters the definition of the momentum, casting it in a manifestly parametrization irrelevant way. 
I.e. this approach uses the Mach's Time Principle primary form $\d Q/\d$(other $Q$) at the primary level in place of the usual approach to physics' notion of velocity.
Note that whilst the velocity, Lagrangian and total Hamiltonian concepts fail when time ceases to be available as a primary concept, the {\sl action} concept does not fail.  
Appendix A's TRiPoD more fully classified which Principles of Dynamics entities survive and which have to be changed, whilst providing the temporally relational replacements for each.

The relational form of the momenta now follow from the temporally relational counterpart of (\ref{Gdyn-momenta})
\be
p^{ij} := \d \, \d\mJ/\d \, \d h_{ij} =   \mM^{ijkl} \d_{\underline{F}}{h}_{kl}/2\,\d{I} \mbox{ } .  
\ee
We then have, as a primary constraint, the quadratic GR Hamiltonian constraint (\ref{Ham}) and, as a secondary constraint, the linear GR momentum constraint (\ref{Mom}).

\noindent The BFO--A equations of motion are  
\be
\d_{\underline{F}}{p}^{ij} = \sqrt{h} \{ R \, h^{ij} -  \, R^{ij} + D^jD^i - h^{ij} D^2 \}\d {I} - 2 \, \d {I}\{ p^{ic}{p_c}^j - p \, p^{ij}/2 \}/\sqrt{h}   \mbox{ } .  
\label{BFO-Evol}
\ee
Finally, the constraint propagation closes as follows: 
\be
\d_{\underline{F}}{\cal M}_i \mbox{ } \mbox{ } = \mbox{ } \mbox{ } D_i\{\{\d{I}\}^2{\cal H}\}/\d I \mbox{ } , 
\label{Mom-dot}
\ee
\be
\d_{\underline{F}} {\cal H} \mbox{ } \mbox{ }  = \mbox{ } \mbox{ } D^a\{\{\d{I}\}^2{\cal M}_a\}/\d {I} \mbox{ } \mbox{ } + \mbox{ } \mbox{ } \d I \, {\cal H} \, p/2\sqrt{h} \mbox{ } .  
\label{Ham-dot}
\ee

\subsection{GR's constraints' Poisson brackets in relational form}\label{GR-PB}

The Poisson bracket between phase-space mixed function--functionals $F := F(x^a; h_{ij}, p^{kl}]$ and $G := G(x^a; h_{ij}, p^{k\ell}]$ is 
\be
\mbox{\bf \{} F \mbox{\bf ,} \, G \mbox{\bf \}} = \int \d^3 z \left\{ \frac{\delta F}{\delta h_{ab}(z)} \frac{\delta G}{\delta p^{ab}(z)}
                                          - \frac{\delta F}{\delta p^{ab}(z)} \frac{\delta G}{\delta h_{ab}(z)} \right\} ~ .
\ee								   
The only slight difference between relational and standard Poisson-brackets formulations of GR is then in the Principles of Dynamics nature of the auxiliary smearing variables
associated with the constraints. 
The TRiPoD indeed entails that these be cyclic differentials rather than Lagrange multipliers.
$\d {K}(x)$ is a scalar integrand cofactor of ${\cal H}$, and $\d L^i(x)$ is a vector integrand cofactor for ${\cal M}_i$.
Furthermore, we assume that the space $\Sigma$ is compact and closed.
The functional form of the smeared constraints is thus as follows.  
\be
(\mathcal H | \d K \, ) := \int d^3 z \, \d K (z) \, \mathcal H (z; h_{ab}, p^{ab}] 
 \,,
\qquad 
(\mathcal M_i | \d L^i )  := \int d^3 z \, \d L^i (z) \,   \mathcal M_i (z; h_{ab}, p^{ab}]  \, .
\ee
The notation we use here reflects that a function-theoretic inner product is being formed.

\noindent The (differential) dynamical evolution of a phase space mixed function--functional $f(x^k; p^{ij}, h_{ij}]$ is given by Poisson brackets with $\int \d A_{\st\so\st\sa\sll}$:
\be
\d f(x^a; p^{ij}, h_{kl}]  = \mbox{\bf \{} f \mbox{\bf ,} \, \mbox{$\int$} \d A_{\st\so\st\sa\sll} \mbox{\bf \}} 
               = \mbox{\bf \{} f \mbox{\bf ,} \, (\mathcal H | \d I) \mbox{\bf \}} 
			   + \mbox{\bf \{} f \mbox{\bf ,} \, (\mathcal M_i | \d F^i) \mbox{\bf \}} = \d I \d f + \pounds_{\sd \underline{F}} f \mbox{ } .  
\ee
Thus the best-matched differential of $f$, $\d_F f$, is the Poisson bracket with $(\mathcal H | \d I)$.

\noindent The algebra of the GR constraints is then as follows (\cite{I93} but with a minor change in the nature of the smearing).
\be
\mbox{\bf \{} (\mathcal M_i | \d L^i \, ) \mbox{\bf ,} \, (\mathcal M_j | \d M^j \, ) \mbox{\bf \}} =  ({\mathcal M}_i | \, [\d L , \d M ]^i )  \, ,
\label{M,M}
\ee
\be
\mbox{\bf \{} (\mathcal H | \d K \, ) \mbox{\bf ,} \, (\mathcal M_i | \d L^i \, ) \mbox{\bf \}} = (\pounds_{\d \underline{L}} {\mathcal H} | \d K \, ) \,     , 
\label{H,M}
\ee
\be 
\mbox{\bf \{} (\mathcal H | \d J \, ) \mbox{\bf ,} \, (\mathcal H | \d K \, )\mbox{\bf \}}  = ( {\mathcal M}_i |  {\d J} \, \overleftrightarrow{\pa}^i {\d K})  \, , 
\label{H,H}
\ee
where $F \overleftrightarrow{\pa}^i G := \{\pa^i F \} G - F \pa^i F$ -- a notation familiar from QFT -- and $[\d L, \d M] = \pounds_{\d \underline{L}}\d M$ is the Lie bracket.  

\mbox{ } 

\noindent Note that this culminates the almost-Dirac procedure with an algebraic closure of the constraints.
Thus indeed classical GR has no Constraint Closure Problem.
\noindent The first two brackets have simple and standard interpretations: that Diff($\Sigma$) is a Lie algebra and ${\cal H}$ is a scalar density respectively.
\noindent The third bracket, however, is harder: making the right-hand side's dependence on the configurational variables explicit, 
\be
( \mathcal M_i h^{ij}|  {\d J} \, \overleftrightarrow{\pa_j} {\d K})  \, , 
\ee
it is manifest that this contains structure functions. 
Moreover, the $h^{ij}$ are in general quite complicated functions of the $h_{ij}$ themselves.
By this, these Poisson brackets form not an algebra but an algebroid.\foo{In the GR literature, Bojowald pointed out that this should be termed an algebroid rather than an algebra 
\cite{BojoBook}. See e.g. \cite{Algebroid} for previous mathematical literature on such algebroids.}
%
Unlike for a rotation transformation, the form a split-spacetime diffeomorphism transformation takes depends on what object it acts upon \cite{T73}.


\noindent Note that henceforth in this paper, we supplant `propagation equations' like (\ref{Mom-dot}, \ref{Ham-dot}) by the Poisson brackets structure formed by the constraints.


\noindent Moreover, the pictorial form for GR's constraints brackets is given a geometrical interpretation in Fig \ref{Teit-Pic-4}. 
In particular, Fig \ref{Teit-Pic-4}.c) illustrates the Refoliation Invariance property of classical GR, by which it avoids the Foliation Dependence Problem.   

{            \begin{figure}[ht]
\centering
\includegraphics[width=0.97\textwidth]{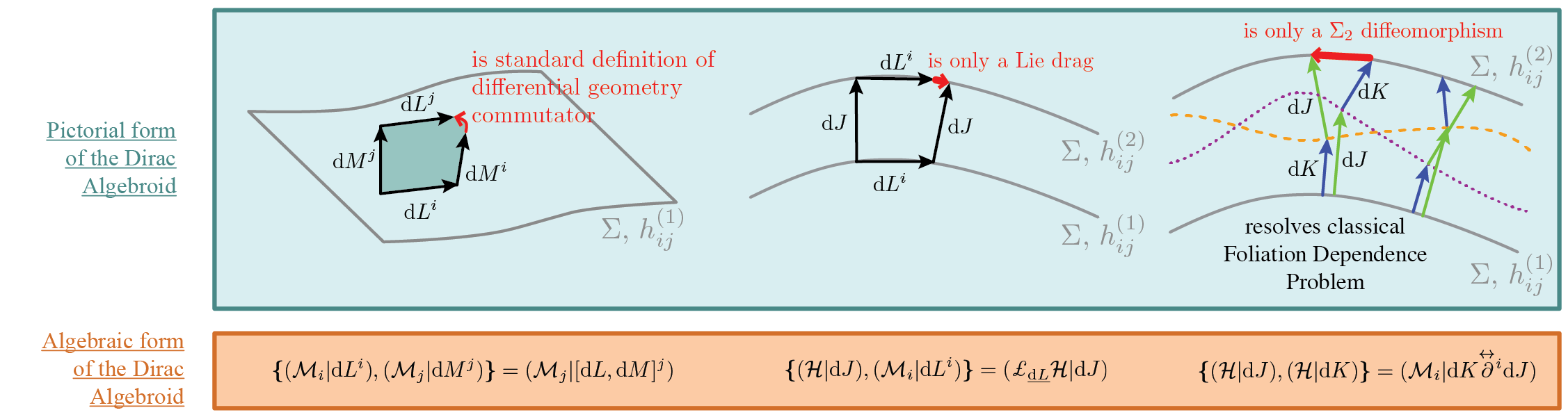} 
\caption[Text der im Bilderverzeichnis auftaucht]{        \footnotesize{Geometrical significance of the form of the Dirac algebroid formed by the constraints of GR.  

\noindent a) ${{\cal M}}_i$ generates stretches {\sl within} $\Sigma$: is just the usual Lie algebra relation in the case of $\fG$ = Diff($\Sigma$).   

\noindent b) Geometrical rendition of ${\cal H}$ is a good Diff($\Sigma$) object.

\noindent c) ${\cal H}$ acts on spatial hypersurface $\Sigma$ by deforming it into another hypersurface (dotted line) \cite{T73, HKT76, Kuchar76}.  
I.e. Teitelboim's Refoliation Invariance construct that holds at the classical level thanks to the form of the Dirac algebroid.  
One considers going between $\Sigma, h^{(1)}_{ij}$ and $\Sigma, h^{(2)}_{ij}$ via 2 different orderings of deformations: 
the first via the dotted hypersurface and the second via the dashed one. 
Then one {\sl only} needs the a stretch within $\Sigma$, $h^{(2)}_{ij}$ to compensate for the non-commutativity of the two deformations involved.} }
\label{Teit-Pic-4} \end{figure}          }

\noindent The next Sec culminates the algebroid strategy for 3 of the facets with classical GR's successful Spacetime Reconstruction from space alone.  
Thus the Spacetime Reconstruction Problem is avoided.

\section{Relativity without Relativity}\label{RWR}

\subsection{BFO--A ans\"{a}tze for first-principles geometrodynamics}

Take the physically irrelevant transformation group $\fG$ to be Diff($\Sigma$), and make the ansatz \cite{RWR, San, Lan} 
\be
\mS^{w,y,a,b} = \int \int \d^3x \sqrt{\sqrt{h}\{a R + b\}} \, \d \ms_{w,y} \mbox{ } .  
\label{trial} 
\ee
Here $\d \ms_{w,y}$ is built out of $\bM_{w,y}$ with components $\mM^{abcd}_{w,y} := \sqrt{h}\{h^{ac}h^{bd} - w h^{ab}h^{cd}\}/y$, and of the usual $\d_{\underline{F}}$.  
The inverse of $\bM_{w,y}$ has components $\mN_{abcd}^{x,y} := y\{  h_{ac}h_{bd} -  x h_{ab}h_{cd}/2 \}/\sqrt{h}$ for  $x := 2w/\{3 w - 1\}$.  
This $x$ has been chosen such that GR is the $w = 1 = x$ case.  
$w = \frac{1}{3}$ is excluded due to non-invertibility, as is $x = \frac{2}{3}$

The conjugate momenta are \cite{FileR} 
\be
p^{ij} = \mM^{ijkl}_{w,y} \d_{\underline{F}} h_{kl} / 2 \, \d I  \mbox{ } . 
\ee
We then have, as a primary constraint, the quadratic constraint 
\be
{\cal H}_{x,y,a,b} := y\{ p^{ab}p_{ab} - x p^{2}/2\}/\sqrt{h} - \sqrt{h}\{a R + b\} = 0 \mbox{ } ,
\ee
and, as a secondary constraint, just the usual ${\cal M}_i$ again \cite{RWR}. 
Thus if one takes Diff($\Sigma$) to be physically meaningless, 
this gives no choice apart from Appendix C's contracted Codazzi embedding equation for the corresponding linear constraint.

Next, the Poisson brackets between the constraints are
\be
\mbox{\bf \{} (\mathcal M_i | \d L^i \, ) \mbox{\bf , } (\mathcal M_j | \d M^j \, ) \mbox{\bf \}} =  (\mathcal M_i | \, [\d L , \d M ]^i )   \,  ,
\label{M-M-2}
\ee
\be
\mbox{\bf \{} (\mathcal H_{x,y,a,b} | \d J \, ) \mbox{\bf , } (\mathcal M_i , \d L^i \, ) \mbox{\bf \}} = (\pounds_{\d \underline{L}} \mathcal H | \d J \, ) \, , 
\label{Htrial-M}
\ee
$$ 
\mbox{\bf \{} (\mathcal H_{x,y,a,b} | \d J \, ) \mbox{\bf , } (\mathcal H_{x,y,a,b} | \d K \, ) \mbox{\bf \}}  = 
- 2 a \, y \, ( D^j p^i\mbox{}_j                                                 +             \{x - 1\}  D_i  p \, | \, {\d J} \, \overleftrightarrow{\pa}^i {\d K}  )
$$
\be
=   a \, y \, ( \mathcal M_i                                                     + 2           \{1 - x\}   D_i p \, | \, {\d J} \, \overleftrightarrow{\pa}^i {\d K}  )
=   a \, y \, ( \mathcal M_i |  {\d J} \, \overleftrightarrow{\pa}^i {\d K}) + 2 \, a \, y     \{1 - x\} ( D_i p \, | \, {\d J} \, \overleftrightarrow{\pa}^i {\d K}  ) \, .
\label{Htrial-Htrial}
\ee
The first two Poisson brackets coincide with the previous Sec's (\ref{M,M}--\ref{H,M}); there is geometrical intuition for this to be the case.  
On the other hand, the evaluation of the last Poisson bracket gives an {\it obstruction term to the algebraic brackets structure},
\be
2 \times a \times y \times \{ 1 - x \} \times ( D_i p  \, | \, {\d J} \, \overleftrightarrow{\pa}^i {\d K}  ) \mbox{ } .  
\ee 
This parallels the brackets algebra obstructing facet of quantum anomalies. 
Moreover, this is not the only facet there is to anomalies; the present example is obviously not a specifically quantum impediment to a perfectly good classical symmetry.  


\subsection{Strongly vanishing options give the three local relativities}

\noindent First consider each of the three options to have this vanish strongly; any of these gives automatic closure and render ${\mathcal H}$ first-class.
We shall see that these correspond to the three qualitatively different kinds of local relativities.  

\mbox{ } 

\noindent {\bf Case 1: General Relativity} Strongly fix the supermetric coefficient to the DeWitt value $x = 1 = w$.
This corresponds to the recovery of \cite{RWR} Einsteinian gravity (GR) with its embeddability into spacetime, whether the usual 
Lorentzian-signature version (-- + + + signature: $a > 0$) 
or its Euclidean counterpart (+ + + + signature: $a < 0$).
The former case possesses locally--Lorentzian relativity, corresponding to a finite maximum propagation speed and null cone. 
(For now, `propagation speed' and `null cone' refer to gravity. 
See Sec \ref{Matter} for how simple matter fields subsequently have to share this finite propagation speed and null cone. 
Thus they share these things with each other, indeed rendering these a universal propagation speed and a corresponding universal null cone.) 
The latter case does not occur physically as is manifested by the existence of a finite propagation speed. 
(Euclidean GR's system of equations is elliptic, corresponding to the 4-manifold having no distinction between time and space and instantaneous action at a distance.)

As regards addressing the classical Spacetime Reconstruction Problem, 

\noindent I) note the dual space-spacetime nature of $h_{ij}$ and $K_{ij}$ due to these being hypersurface tensors.   
\noindent In the BFO--A 

\noindent version of the sandwich construct, the data is $h_{ab}$ and $\d h_{ab}$ (Machian form of the Thin Sandwich Problem).
Then $K_{ab}$ comes out as
\beq
K_{ab} = \d_{\underline{F}} h_{ab}/2 \, \d I \mbox{ } . \mbox{ } \mbox{ I.e. it is of the Machian relational form  d(change)/ d(other change)} \mbox{ } . 
\eeq
Furthermore all the geometrical change is given the opportunity to contribute to the $\d I$ that each individual change is compared to: a STLRC concept . 
This is contrasted with other formulations' versions in Appendix B.  

\noindent II) Both the Euclidean and Lorentzian cases entail an embedding into a manifold of dimension 1 higher by the ${\cal H}$ now taking the form of the double contraction 
(\ref{G00}) of  Appendix C's Gauss embedding equation (\ref{GE}).  
Thus it matches the contraction (\ref{G0i}) of the Codazzi embedding equation (\ref{G0i}) that is the GR momentum constraint. 
This is so as to form a pair of embedding equations \cite{T73} that are 4 of the 10 components of the Einstein field equations  (\ref{G00H}, \ref{G0iM}).
The equations of motion turn out to be a linear combination of the Ricci embedding equation (\ref{RE}), of the contracted Gauss embedding equation and of the metric times further 
contractions. 
By this, they form the temporally relational version of the remaining 6 Einstein field equations (\ref{Gab2}).
The form of the constraints that comes out of the procedure is consistent with the $0\mu$ components of the 4-$d$ Einstein tensor, and the evolution equation can be recast into the 
form of the $ij$ components of the Einstein tensor.

Here we {\sl recover} equations and make a {\sl meaningful grouping} of them. 
This is as opposed to the opposite-sense decomposition by embedding equations alluded to in the Introduction, or the assumption of embeddability into spacetime of \cite{T73, HKT76}.  
This set of equations thus gives spacetime (Lorentzian or Euclidean), at least within a localized sandcastle-shaped region in the Lorentzian case. 
This is further reinforced by suitable Analysis (GR Cauchy Problem).  
These equations are valid in two local senses (in space and in time).
I.e. for a sandcastle-shaped piece of the domain of dependence  [Fig \ref{Sandcastle}] has well-posedness suitable to hyperbolic-type equations with Cauchy data. 
See \cite{LerayBruhat, Wald} for results on this based on older approaches, and \cite{HE, HKM-CBY80, Rendall} for more modern ones. 

{            \begin{figure}[ht]
\centering
\includegraphics[width=0.55\textwidth]{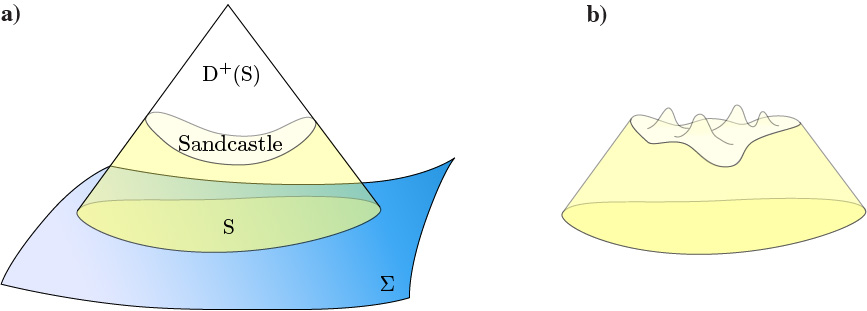} 
\caption[Text der im Bilderverzeichnis auftaucht]{        \footnotesize{ a) The domain of dependence of a piece $\mS$ of a spatial hypersurface $\Sigma$, $\mD^+(\mS)$, is the portion 
of spacetime that is controlled solely by the physical data on $S$.  
Points on $\Sigma$ outside $\mS$ are not able to causally communicate with $\mD^+(\mS)$.  
Within this is shaded an example of `sandcastle-shaped' region for which the GR Cauchy Problem results could be expected to hold.
This would be more strictly sandcastle-shaped if the wedge it is a piece of had one angular dimension suppressed; however, it has two angular dimensions suppressed. 

\noindent b) Its surface may in fact be wiggly since different parts of the data differing in how far they can be evolved.]} }
\label{Sandcastle}\end{figure}            }

The BFO--A action results from this first strong fixing, and can be repackaged as, firstly, the A action (\ref{S-A}).  
Secondly, it can be cast as the Einstein--Hilbert action for the sandcastle-shaped region using just the one embedding equation rather than the three used at the level of the equations of motion.  

\mbox{ } 

\noindent {\bf Case 2: Galileo--Riemann Geometrostatics}. 
Strongly fix $y = 0$ \cite{Lan}.  
This involves the trial quadratic constraint ceasing to contain a kinetic term.  
As far as we know, this theory was first pointed out by Teitelboim \cite{T80}.
It is to be interpreted as non-dynamical: a {\sl geometrostatics}, which additionally possesses \cite{Lan} Galilean relativity. 
I.e. the well-known limit of Lorentzian relativity as the maximum propagation speed $c_{\sg\sr\sa\sv}$ becomes infinite, in which the null cones become squashed into planes.  
Moreover the underlying notion of space is in general not just $\mathbb{R}^3$ but rather any fixed Riemannian 3-space.\footnote{In the absense of matter, 
it has to be a constant Ricci curvature space by the statement of the constraint, but this restriction is lost upon inclusion of matter in Sec \ref{Matter} \label{fooref}.}
Thus, this case does not just cover Galilean relativity in its original context as obeyed by Newton's mechanics. 
More broadly, it covers its `Galileo--Riemann' generalization to the possible worlds envisaged by Riemann that have in general non-flat Riemannian spatial geometry that is, 
nevertheless, a {\sl fixed} spatial geometry.

Note that if one insists that one's {\sl action} must be built from first principles, this Galilean relativity option is not possible.
This corresponds at the most primary level to relationalism precluding a geometrodynamics in which the geometry indeed undergoes nontrivial dynamics.
On the other hand, if one attributes primary significance to the algebraic structure of the constraints, this option is allowed in both such a restriction of the relational approach 
and in Teitelboim's previous program \cite{T80}.

Since this model has no dynamics of geometry, any supporting Analysis involves equations containing only spatial derivatives.
These are second-order, so it is an elliptic-type problem; by footnote \ref{fooref}, it concerns whether constant Ricci curvature spaces exist for a given metric. 
Moreover, this is unrelated to the space-time structure of the theory, which is just a stack of copies of the same 3-space (assuming that this exists).  
I.e. a fixed foliation the leaves of which are all equal.  
Given the global existence of the geometry that constitutes one copy of the leaf in question, nothing stops this simple construct extending ad infinitum in the time direction.  
Finally, infinite propagation speed has rendered the notion of domain of dependence trivial.  

\mbox{ } 

\noindent {\bf Case 3: Strong Gravity/BKL regime}. 
Now fix $a = 0$ \cite{San}. 
This involves the potential ceasing to contain a Ricci scalar, which corresponds to strong gravity \cite{Isham76}, which possesses Carrollian relativity \cite{LL65-BLL68}.
This is the less known opposite limit to Galilean relativity of Lorentzian relativity.  
Here the maximum propagation speed $c_{\sg\sr\sa\sv}$ goes to zero, so that the null cones become squeezed into lines,  so that each point can only communicate with its own worldline.
Carrollian relativity is named after Lewis Carroll, for the Red Queen's musing {\it ``Now, here, you see, it takes all the running you can do, to keep in the same place"} \cite{Carroll}.

Strong gravity is most well known in the case that is a a strong-coupled limit of GR (so $w = 1 = x$) \cite{Isham76}. 
This is widely believed to apply to the primordial universe near a Belinsky--Khalatnikov--Lifshitz singularity \cite{BKLpaper}.
However, removing the above obstruction term does not require thus fixing the supermetric coefficient, $w$. 
The other values of $w$ now correspond to other consistent geometrodynamical theories thus discovered \cite{San}.  
This then turns out to be interpretable as strong-coupled limits of scalar--tensor theories that likewise apply near singularities in those theories.  
Clearly from each worldline only being able to communicate with itself, elsewhere than near singularities these geometrodynamics theories are {\sl completely unrealistic}.

Also not that that \cite{Henneaux79} can be readily interpreted to affirm that the hypersurface derivative--best matched derivative maintains 4-space--3-space `dual 
nationality', except that the nature of the 4-objects is distinct.
Henneaux \cite{Henneaux79} and Teitelboim \cite{T80} followed this up by working out the strong gravity analogue of 'hypersurfaces in spacetime' geometry.
Strong gravity has a degenerate-signature manifold (0 + + +) for its space-time structure \cite{Henneaux79}. 
Thus it serves as an example of how the above-mentioned `dual nationality' is not an exclusive property of spacetime and its Euclidean counterpart.

Additionally, strong gravity's space-time construction can be supported with a simpler sort of Analysis -- now for o.d.e's spatial point by spatial point as per \cite{SG-Dyn}.   
Then e.g. Rendall goes as far as describing a widely applicable global existence theorem for such o.d.e's \cite{Rendall}.  
Another consequence of the Analysis being spatial point by spatial point is that the domain of dependence wedge is replaced by a `solid cylinder'.  
This is composed of the worldlines of the points within a `cross-section' region $\mS$ each line within which is its own domain of dependence.
Though some of these lines may extend further than others due to better fortunes with existence results.  
Thus this `cylinder' may in practise be wiggly, much as Fig 3's  `sandcastle' is. 
%

\mbox{ }

\noindent Note that the zero and infinite propagation speed cases additionally greatly simplify the algebraic structure formed by the constraints.
This is because these are {\sl factors in common} with the momentum constraint arising in the bracket of two trial-Hamiltonian constraints.
By this, the momentum constraint is no longer an integrability of the trial-Hamiltonian constraint.  
Additionally, the algebraic structure no longer involves any structure functions, so that it is a bona fide algebra rather than an algebroid.
Namely, 
\be
\mbox{\bf \{} (\mathcal M_i | \d L^i \, ) \mbox{\bf ,} \, (\mathcal M_j | \d M^j \, ) \mbox{\bf \}} =  (\mathcal M_i | \, [\d L , \d M ]^i )   \,  ,
\label{M-M-3}
\ee
\be
\mbox{\bf \{} (\mathcal H | \d J \, )     \mbox{\bf ,} \, (\mathcal M_i | \d L^i \, ) \mbox{\bf \}} = (\pounds_{\d \underline{L}} \mathcal H | \d J \, ) \, , 
\label{H-M-3}
\ee
\be 
\mbox{\bf \{} (\mathcal H | \d J \, )     \mbox{\bf ,} \, (\mathcal H | \d K \, ) \mbox{\bf \}}  = 0 \, ,
\label{H-H-Abelian}
\ee
is obeyed by both the Carroll and the Galileo--Riemann case. 
These correspond to entirely opposite representations of the object ${\cal H}$: pure-potential and pure-kinematical plus $\Lambda$-term cases respectively. 
Isham \cite{Isham76} first unveiled this algebra in the Carroll case, 
whilst Teitelboim \cite{T80} additionally pointed out that it holds in what we later identified as the Galileo--Riemann case. 

Also note that the three possible strong ways of evading the obstruction term are the trichotomy of zero, finite and infinite propagation speed relativities.  
\noindent Now localized relativity follows from closure of the algebraic structure of the constraints rather than being postulated a priori as half of a two-way switch \cite{Rindler} (the other half 
being Galilean relativity) as it was historically by Einstein.  
This result parallels the brackets algebra obstructing facet of quantum anomalies.  
In this sense, the three types of localized relativity arise in the same manner as the much-vaunted critical dimensions 26 and 10 for bosonic strings and superstrings respectively. 
This is in the sense that all of these are picked out as strong impositions that close a brackets algebra. 
Moreover, the present paper's case involves     the {\sl classical} Poisson brackets closure as a Dirac    algebroid (\ref{M,M}--\ref{H,H}) 
                                                                                  or the simpler bona fide algebra (\ref{M-M-3}--\ref{H-H-Abelian}).
This is as opposed to the String Theory case's                {\sl quantum} commutator closure as a       Virasoro algebra or its supersymmetric counterpart \cite{GSW}.
We find the present paper's case of particular physical and philosophical interest because removing the obstruction here involves picking a coefficient to take values that 
mathematize the decision of which type of localized relativity to adopt.
Moreover, if one adopts the physical choice -- locally Lorentzian relativity. 
This comes hand in hand with {\sl deducing} embeddability into a spacetime perspective that has long been known to be widely insightful \cite{HE, Wald}. 
This is as opposed to requiring fixing a coefficient to take a surprising value, such as the dimension of spacetime being 26 or 10.  
Such surprises subsequently require a lot more explanation (hiding extra dimensions by e.g. compactifications or braneworld-type warpings).

\subsection{The weakly-vanishing option gives CMC slicing}\label{3.3}

The proper Poisson brackets form of the d{\it A}-Dirac analysis makes it clear that $D_ip = 0$ is the only form of condition that ensures the weakly vanishing option. 
Integrating up, one possibility this gives is 
\beq
{\cal D} := p - \sqrt{h}c = 0
\label{CMC}
\eeq
for $c$ spatially constant.  
This is the conventional notion of constant mean curvature (CMC).
It is called CMC because, via the momentum--extrinsic curvature relation (\ref{Gdyn-momenta}), it implies that the mean curvature (proportional to the trace of the extrinsic curvature, $K$), is constant.  
It has distinguished maximal subcase ($c = 0$), which however proves to be too restrictive for the combination of consistency and physical plausibility (see below and Sec \ref{3.6}).
A more general possibility is to use a particular kind of functional in place of (\ref{CMC})'s straightforward constant: 
\beq
{\cal D}_1 := p - \sqrt{h} \big\langle p/\sqrt{h} \big\rangle = 0 \mbox{ } .
\eeq
Both to explain the notation and to wrap up, let $\pi := p/\sqrt{h}$,  $\langle A \rangle := {\int d^3 x  \sqrt h \, A }\big/{\int d^3 x \sqrt h}$ 
be the standard definition of average over all of space. 
Since we took this to be compact without boundary, this makes sense. 
This is also written for an {\sl undensitized} object $A$, and $\overline{A} := A - \langle A \rangle$, the `unnormalized inhomogeneity quantifier'.
Then
\beq
{\cal D}_1 = \sqrt{h}\overline{\pi} = 0 \mbox{ } .
\eeq

Note that the second possibility above is zero for spatially-homogeneous universes. 
In fact, for these the $D_i$ operation annihilates everything in this model, so that the weak avoidance of the obstruction term is automatic.
Additionally, it is well-known that there is no ${\cal M}_i$ in this case, so the constraint algebra ends up being just ${\cal H}_{x,y,a,b}$ commuting with itself. 
Generalized minisuperspace is {\sl not} restricted by consistency.of the algebraic structure of the constraints.  
This is very similar in content to strong gravity. 
($R$ is a spatial constant.  
Thus $a R + b$ is just like $0 + b^{\prime}$ for a new constant $b^{\prime}$).  
Whilst closely resembling strong gravity, this minisuperspace option can be used as a back-cloth for locally Lorentzian physics viewed as perturbations that negligibly disturb the 
gravitational sector. 
All the values of $w$ other than 1 can then be interpreted as the consistent minisuperspaces belonging to scalar-tensor theories \cite{San}.  
This rather suggests investigating the CMC route (and all its variants below and in Sec \ref{3.6}) round the obstruction term for slightly inhomogenous cosmology.

For the rest of this Section, we concentrate on the CMC case for specifically-inhomogeneous GR.
Let us first consider ${\cal D}$ to be a new constraint.  
Smear it with some differential scalar $\d\sigma(z)$:  
%
%
$(\mathcal D | \d\sigma)  := \int d^3 z \, \d\sigma (z) \, \mathcal D (z)$. 
We now need to evaluate its Poisson brackets with the whole set of constraints hitherto found. 
\be
\mbox{\bf\{} (\mathcal D |  \d\sigma \, ) \mbox{\bf ,} \, (\mathcal D |  \d\rho \, ) \mbox{\bf \}}  = 0 \, ,
\ee
\be
\mbox{\bf\{} (\mathcal D |  \d\sigma  ) \mbox{\bf ,} \, (\mathcal M_i |  \d L^i \, )\mbox{\bf\}}   =  ( \pounds_{\d \underline{L}} \mathcal D | \d\sigma \, )  \approx 0 \,  ,
\ee
which are uneventful.  
I.e. ${\cal D}$ forms an Abelian subalgebra since it acts simply enough on the smearing to produce a $\d\sigma\,\d\rho -\d\rho\,\d\sigma = 0$ factor in the right-hand side
and ${\cal D}$ is a scalar density as regards the 3-diffeomorphisms (just as ${\cal H}_{x,y,a,b}$ is). 
By thus closing so far no second-classness has appeared in the algebraic structure of the Poisson brackets.
This changes with the next bracket to be evaluated:
\beq
\mbox{\bf\{} (\mathcal H_{x,y,a,b} |  \d J  \, ) \mbox{\bf ,} \, (\mathcal D|  \d\sigma \, ) \mbox{\bf \}}    
= ({\cal L} \, \d J \, | \d\sigma) - \mbox{$\frac{3}{2}$}({\cal H}_{x,y,a,b}| \d \sigma \, \d J)   \mbox{ } . 
\eeq
Here, ${\cal L} \, \d J = 0$ is a differential of the instant fixing equation type of specifier equation. 
I.e. it is an example of the fourth `specifier equation' option in the d{\it A}-Dirac procedure and the obvious parallel of the more well-known if not TRiPoD notion of lapse-fixing equation.
The corresponding second-order linear differential operator takes the form  
\begin{equation}
\mathcal L := 2 \, a \, D^2 - \{ 2 \, a \, R + 3 \, b \} - y \, \left\{ {\textstyle \frac 3 2} x - 1 \right\} \, c \, p/\sqrt{h}  \,,
\end{equation}

Furthermore, it is a temporally relational recasting and generalization (by the presence of coefficients $a, x, y$) of a particular well-known lapse-fixing equation. 
That is the one that fixes (at least for a local time-interval ) a CMC-foliation. 
Also, by such a right-hand side term being present in the brackets structure, $\mathcal D$ has knocked ${\cal H}$ off its `so far first class constraint' perch. 
Thus if $\mathcal D$ is regarded as a constraint, it and ${\cal H}$ are second-class constraints. 
Thus the degrees of freedom count is now $12-3\times 2-2 \times 1 = 4$. 
I.e. overall unaffected by there being a new constraint but its presence causing the status of an already-existing constraint to change. 
The specifier equation itself, if regarded as an entity to propagate, would just form a tower of conditions on an equally-increasing number of freedoms.

Let us next consider $\mathcal D_1$ to be a new constraint. 
The first two Poisson brackets are as before with $\mathcal D \rightarrow \mathcal D_1$ since the reasons for the forms of these are preserved. 
On the other hand,
\begin{equation} \label{EQUATIONpatchNO1}
\mbox{\bf \{}  ( \mathcal{H}_{x,y,a,b} | {\rm d} J ) \mbox{\bf ,} \,  (\mathcal D_1 | {\rm d} \sigma)  \mbox{\bf \} }
= (  \mathcal L_1 {\rm d} J  | \overline{ {\rm d} \sigma }) \,,
\end{equation}
(notice the bar over ${\rm d} \sigma$), where now
\begin{equation}
\mathcal L_1  := 2 \, a \, D^2 - \{ 2 \, a \, R + 3 \, b \} - y \, \left\{ {\textstyle \frac 3 2} x - 1 \right\} \, \langle \pi \rangle \, p \, .
\end{equation}
Then the differential of the instant fixing equation can be written (for $a \neq 0$ and ${\cal O} := {\cal L}_1/2a$)
\begin{equation}\label{EQUATIONpatchNO2}
\overline{{\cal O}\d J} = 0 \mbox{ } . 
\end{equation}
\noindent Eq. (\ref{EQUATIONpatchNO2}) is of the form $D^2 {\rm d} J - M_{x,y,a,b} \, {\rm d} J = \text{\it const.}$. 
The term $M_{x,y,a,b} :=  \, R + {\textstyle \frac{3 \, b}{2 \,a} }  + {\textstyle \frac y {4 \, a}} \, \left\{ 3 \, x - 2 \right\} \, \langle \pi \rangle p$  
is positive for a choice of parameters $x,y,a,b$. 
In fact we can rewrite it by decomposing the momenta into traceless part $p_{\rm \tT}^{ab}$ and trace part $\langle p \rangle$,
\begin{equation}
p^{ab} = p_{\rm \tT}^{ab} + {\textstyle \frac 1 3} \langle \pi \rangle \, \sqrt h \, h^{ab} \,, ~~~~ \Rightarrow ~~~~
p^{ab} p_{ab} = p_{\rm \tT}^{ab} p^{\rm \tT}_{ab}  + \langle \pi \rangle^2 \, h \,,
\end{equation}
and using the Hamiltonian constraint $\mathcal H_{x,y,a,b} \approx 0$ to eliminate $R$ and the CMC constraint $\mathcal D_1 \approx 0$ to
eliminate $p$:
\begin{equation}
M_{x,y,a,b}  \approx   {\textstyle \frac y a} \frac{p_{\rm \tT}^{ab} p^{\rm \tT}_{ab}}{h} + {\textstyle \frac{3 \, b}{2 \,a} }  + 
{\textstyle \frac y {4 \, a}} \, \left\{ 3 \, x + 2 \right\} \, \langle \pi \rangle^2 \,, 
\end{equation}
we see that $M_{x,y,a,b} \geq 0$ if $b \geq 0$, $y\geq 0$ and $a \geq 0$. 
Let us stick to this case: we can now prove that the differential of the instant fixing equation $D^2 {\rm d} J - M_{x,y,a,b} \, {\rm d} J = \text{\it const.}$ has a one-parameter set 
of solutions. In fact the associated \emph{homogeneous equation:}
\begin{equation}
D^2 {\rm d} J - M_{x,y,a,b} \, {\rm d} J = 0 \,,
\end{equation}
admits only the  trivial solution ${\rm d} J = 0$.  
In fact if there was a nonzero solution ${{\rm d} J}_0$, since the manifold is compact, ${{\rm d} J}_0$ would admit a maximum at some point $x_1$ and a minimum at some other point $x_2$. 
At those points $D^2 {{\rm d} J}_0(x_1) < 0$ and $D^2 {{\rm d} J}_0(x_2) > 0$ which imply via eq. (\ref{EQUATIONpatchNO1}) that $M_{x,y,a,b} (x_1)\, {{\rm d} J}_0 (x_1) <0 $ and 
$M_{x,y,a,b} (x_2)\, {{\rm d} J}_0(x_2) > 0$. 
Since $M_{x,y,a,b} $ is positive this implies that ${{\rm d} J}_0(x_1) < 0$ and ${{\rm d} J}_0(x_2) > 0$. 
But then we have a negative maximum and a positive minimum, which is a contradiction.
Therefore, at least in the case $b \geq 0$, $y \geq 0$, $a \geq 0$, the homogeneous equation admits only the zero solution. 
This means that the operator $D^2 - M_{x,y,a,b}$ has an empty kernel, and is therefore invertible. 
This then implies that the inhomogeneous  equation admits a unique solution $D^2 {\rm d} J - M_{x,y,a,b} \, {\rm d} J = \text{\it const.}$.

The equation we have to solve, though, is not strictly inhomogeneous. 
It depends on a single global linear combination of the values of ${\rm d} J$ through the right hand side term $ \left< M_{x,y,a,b} \d J \right>$. 
This makes it a \emph{reducible} system. In other terms, the equation is invariant under rescalings ${\rm d} J  \rightarrow \kappa \, {\rm d} J$ where $\kappa$ is a spatial constant. 
The conclusion is that, in the case $b \geq 0$, $y\geq 0$, $a \geq 0$, Eq. (\ref{EQUATIONpatchNO2}) admits a one-parameter family of solutions.
I.e. one for each possible normalization of  ${\rm d} J$). 
All of these solutions are physically equivalent, as they correspond to an unimportant constant rescaling of the differential of the instant, or a redefinition of the unit of time.

\mbox{ } 

\noindent We conclude that various further interpretations are possible if one reformulates this means of attaining consistency (see Sec \ref{3.6}). 
Thus the zero-finite-infinite trichotomy of Carrollian, Lorentzian and Galilean relativities comes out 
hand in hand with an unexpected fourth almost-equal partner: a CMC-sliced theory.\footnote{`Perfectly homogeneous' 
is also a mathematically-valid fifth option, but that is obviously overruled by almost every scientific observation ever made.)}
%
The obstruction term's factors represent a joint packaging not seen before in Theoretical Physics. 
\cite{ABFKO}'s variational principle adds weight to York's initial value problem formulation.  
Additionally, the 4-factor equation to his time being an internal time with some of the less garish trappings of absolute time sitting hidden within GR. 
However, Isham., Kucha\v{r} and one of us (E.A.) have levelled other arguments against York time \cite{Kuchar92, I93, APOT2}.

\subsection{Metrodynamics assumed}

On the other hand, if we take $\fG$ = id, 
\be
\mT = ||\d \bh||_{\sbM}\mbox{}^2 \mbox{ } . 
\ee
Then we have just a primary ${\cal H}_{x,y,a,b}$ constraint to start off with. 
Its propagation is as before, except that, firstly, it involves a plain $\d$ and not a $\d_{\underline{F}}$.  
Secondly, we {\sl not} have an initial right to a priori `parcel out' an ${\cal M}_i$, so we are to use the first form of the right-hand side of (\ref{Htrial-Htrial}).  
Instead, at the level of the Poisson brackets of the constraints, 
\begin{equation}
\mbox{\bf \{} ( \mathcal H_{x,y,a,b} | \d J ) \mbox{\bf ,} \, ( \mathcal H_{x,y,a,b} | \d K ) \mbox{\bf \}}
=  a y \, (  - 2 D_j p^j\mbox{}_i + 2\{1 - x\} D_ip | \d J \overleftrightarrow{\pa}^i \d K ) \mbox{ } . 
\label{ha}
\end{equation}
Thus we define
\beq
{\cal S}^x_i :=  2\{ -D_j p^j\mbox{}_i + \{1 - x\}D_i p\}
\label{Daenerys} 
\eeq 
as the preliminary secondary constraint entity arising from this Poisson bracket.  
For now we smear this with some differential vector $\d \xi^a$.  
(\ref{Daenerys}) is present unless one of $a = 0$ or $y = 0$ holds, in which case the above right-hand side strongly vanishes.

If $x = 1$ holds, then ${\cal S}^x_i$ collapses to  ${\cal M}_i$ the generator of diffeomorphisms and therefore the main case 1) of RWR's working is recovered.
In fact, the preceding occurs as one of two special cases that permit the algebraic structure of the constraints to close weakly. 
\be
\mbox{\bf \{} ( {\cal S} _i | \pa \xi^i ) \mbox{\bf ,} \, ( {\cal S}_j | \d \chi^j ) \mbox{\bf \} }  =  
(   -2 \, D_j {p^j}_i  + 2\{1 - x\}\{3x -  2 \} D_i p \, | \, [\d \xi, \d \chi]^i) \, .
\label{lf}
\ee
Then comparing (\ref{ha}) and (\ref{lf}), the algebraic structure of the constraints closes on itself only if $\{1 - x\}\{3x -  2 \} = 0$, so $x = 1$ or $x = \frac{2}{3}$.

Moreover, the latter special choice of $x$ also has a clear geometric meaning. 
$$
\mathcal U_i := \mathcal S_i^{\mbox{\scriptsize$\frac 2 3$}} = -2 \left\{ D_j p^{j}\mbox{}_i  - \mbox{$\frac 1 3$} \, D_i p \right\} \,,
$$
is the generator of {\it unit-determinant diffeomorphisms}, i.e. diffeomorphisms that leave the local volume element $\sqrt h$ invariant.
This corresponds to taking the traceless part of the momentum
\begin{equation}
 ( \mathcal U_a | \d \xi^a  )  = - 2 \int   \d\xi_a D_b \{ p^{ab}  - \mbox{$\frac 1 3$} \, h^{ab} \, p \} \,    
                               = \int  p^{ab}  \left\{ \pounds_{\d \underline{\xi}} h_{ab} - \mbox{$\frac 2 3$} \, h_{ab}D_c \d \xi^c  \right\} 
                               = \int  p^{ab}  \, h^{\frac 1 3} \, \pounds_{\d \underline{\xi}}\{ h^{-\frac 1 3} h_{ab}\} = \int  p^{ab}  (\mathbb{L} \d\xi)_{ab} \,.  
\end{equation}
Here $h^{-\frac 1 3} h_{ab}$ is the unimodular metric and $(\mathbb{L}\d\xi)_{ab} := D_a \d \xi_b + D_b \d \xi_a - \mbox{$\frac 2 3$} \, h_{ab} D_c \d \xi^c $ is the 
{\it conformal Killing form} acting upon $\d \xi^c$.
Adopting $x = \frac{2}{3}$ corresponds at the level of Riem($\Sigma$) to picking the degenerate (null signature) supermetric. 
This degeneracy means that this option has no underlying relational action.  
Finally, 
\begin{equation}
\mbox{\bf \{} ( \mathcal U_a | \d \xi^a  ) \mbox{\bf , } ( p | \d \varphi ) \mbox{\bf \} } = 0  \,,
\end{equation}
\begin{equation}
\mbox{\bf \{} (   \mathcal H_{x,y,a,b} | \d J  ) \mbox{\bf , } ( D_a p | \d \xi^a )  \mbox{\bf \} }  = 
\mbox{$\frac{3}{2}$}\left( \mathcal H_{x,y,a,b} | \, \d J \, D_a\d\xi^a \right) +   
\left(  \{ a R + 3 \, b -  D^2  \} {\rm d} J \, | \, D_c {\rm d} \xi^c   \right) \mbox{ } .  
\end{equation}
Thus $x = \frac{2}{3}$ gives a derivative of the instant fixing equation subcase of d{\it A}-Dirac procedure's specifier equation, 
\begin{equation}
 D_i \left\{\{ a R - 6 \, \Lambda\} \d J  - D^2 \d J \right\} = 0 \mbox{ } \mbox{ } \Rightarrow \mbox{ } \mbox{ }  
 \{ a \, R - 6 \, \Lambda\} \d J  - D^2 \d J  = \langle   \{ a R - 6 \, \Lambda\} \d \chi  \rangle \mbox{ } .  
\end{equation}
This is pretty much the equation from Shape Dynamics, i.e. without the York time but with the constant term at the right hand side that makes it solvable anyway. 

\noindent This theory's exact meaning remains unknown, but it exemplifies a theory that lies somewhere between a metrodynamics and a geometrodynamics. 
Let us use the terms $U$-diffeomorphism, $U$-geometry and $U$-superspace for this theory's counterparts of the entities in the study of geometrodynamics.  
$U$-diffeomorphisms have 2 degrees of freedom per space point, leaving $U$-geometry with 4.

\mbox{ }  

\noindent If $x \neq 1$ or $\frac{2}{3}$, however, then the outcome of $\mbox{\bf\{} ({\cal H}|\d J) \mbox{\bf ,} \, {(\cal S}^x_i|\d \xi^i) \mbox{\bf\}}$ decomposes into the following. 

\noindent 1) ${\cal M}_i$ as a new constraint distinguished by its own Poisson brackets closing at the tertiary level. 

\noindent 2) A functional of another new constraint whose own brackets compromise the first-class status of ${\cal H}_{x,y,a,b}$ and produce at the tertiary level a specifier equation.
That is the generalized CMC differential of the instant fixing equation.
This decomposition is clearly within the ethos of the d{\it A}-Dirac procedure having four outcomes that can occur in any combination at any level. 
Applying this decomposition requires a matching reassignment of smearing variables: each part of the decomposition requires its own smearing variable. 
Moreover, it also choose a simpler functional form resident within one of the two additive parts rather than the whole of both additive parts. 
I.e., we choose a scalar ${\cal D}$ (or ${\cal D}_1$) and then smear that with a differential scalar $\d\sigma$ (or $\overline{\d \sigma}$). 
This is to be contrasted with choosing $D_i{\cal D} (= D_i{\cal D}_1)$ and then smearing that with some differential vector.

By the above analysis, we have determined the forced return to the preceding section's workings in cases 1) and 4). 
I.e. metrodynamics initially assumed in fact entails no new choices for these cases, the d{\it A}-Dirac procedure bringing in the same constraints that geometrodynamics possesses ab initio.
\noindent In the GR case, having an $aR$ term in the potential gives no choice but contracted Codazzi embedding equation.  
This is via the constraint ${\cal M}_i$ associated with the Diff($\Sigma$) transformations arising as an integrability of any ${\cal H}_{\st\sr\si\sa\sll}$ with such a potential.

On the other hand, there is no need for a ${\cal S}^x_i$ to arise at all if one takes a strong equality $a = 0$ or $y = 0$ way out of the bracket of two trial ${\cal H}$'s.
So in cases 2) and 3), if one does not presuppose geometrodynamics and thus find the associated momentum constraint, 
one does not find that the momentum constraint is enforced none the less as an integrability of the Hamiltonian constraint. 
The algebraic structure now formed by the constraints is now just the Abelian algebra 
\be
\mbox{\bf \{}({\cal H}|\d J) \mbox{\bf ,} \, ({\cal H}|\d K)\mbox{\bf \}} = 0 \mbox{ } .  
\label{Abel}
\ee
This is a distinct 5-degrees-of-freedom strong gravity with Carrollian relativity, or a Riemannian metrostatics with Galilean relativity.
(That is 5 degrees of freedom from 6 $\times$ 2 initially -- 1 $\times$ 2 from ${\cal H}_{\st\sr\si\sa\sll}$ leaving 5 $\times$ 2 phase space degrees of freedom.)
5-degree-of-freedom strong gravity with the GR value $x = 1 = w$, previously considered in e.g. \cite{5-dof},  
and 5-degree-of-freedom Strong Gravity for other values of $x$ or $w$ was discovered in \cite{San}.  
Again, whilst cases 2) and 3) share the same algebra but each involves a diametrically opposite representations of the ${\cal H}$ object itself 
(the mostly kinetic term ${\cal H}_{0,b,x,y}$ and the zero-kinetic term ${\cal H}_{a,b,x,0}$). 

Finally, an unexpected $U$-geometrodynamics has now been found as a fifth distinguished outcome in the case of metrodynamics, rather than geometrodynamics, assumed.  
This option is however precluded if a relational action is demanded.
That gives extra value to the venture of presupposing less features in common with GR's own geometrodynamics.

\subsection{Discover-and-encode approach to Physics}\label{DAE}

At the classical level, this amounts to trying out a $\fG$, finding it gives further integrability conditions that enlarge $\fG$ and then deciding to start afresh with this enlarged 
$\fG$ as one's better-informed first principles.  


\noindent Example 1) For instance, if one insists on finite propagation speed and contemplates metrodynamics ($\fG$ = id), one is forced to philosophize differently in terms of 
geometrodynamics [$\fG$ = Diff($\Sigma$) enlargement of $\fG$ = id]  or $U$-geometrodynamics.  

\noindent Examples 2) and 3) One might likewise take a discover-and-encode attitude to using the fourth factor that follows from the presumption of geometrodynamics.
[$\fG$ = Diff($\Sigma$), starting afresh with the semidirect products \cite{I84} $\fG$ =  Conf($\Sigma$) $\rtimes$ Diff($\Sigma$) \cite{ABFO} or 
$\fG$ =  VPConf($\Sigma$) $\rtimes$ Diff($\Sigma$) 
\cite{ABFKO} enlargements of $\fG$ = Diff($\Sigma$) that entail some form of conformogeometrodynamics, as outlined in the next SSec.]

\noindent It may also happen that an initially posited $\fG$ runs into inconsistencies at the level of the algebraic structure of the quantum commutator brackets.

\subsection{Conformogeometrodynamics assumed}\label{3.6}

\noindent We keep our account of this brief since the auxiliaries now used need to be more complicated than the cyclic differentials employed in the present paper.
I.e. both an auxiliary coordinate and its differential now appear in the relational action.
The action now encodes a conformalized ${\cal H}_{x,y,a,b}$ that is a generalization of the Lichnerowicz--York equation \cite{York7172, York73}.

The conformalized form of Hamiltonian constraint is
\be
y \,  p^{\sT ij} p^{\sT}_{ij} + \hat\phi^{12}p^2\{2 - 3x\}/6  = ah \, \hat \phi^8 \{ R - 8 D^2\hat\phi/\hat\phi \} + bh\hat\phi^{12} \,. 
\label{LYeq}
\ee
One can then regard the Hamiltonian constraint as a gauge fixing that breaks the conformal covariance by selecting a particular global volume preserving 
$\hat\phi = \phi/\{\int_{\Sigma}\d^3x \sqrt{h}\phi^6\}^{1/6}$.

Note that a version of ${\cal D}$ that contains conformalized variables now arises directly from variation.  
So does the specifier equation (\cite{ABFKO}'s lapse-fixing equation recast as a velocity of the instant fixing equation).

This consistent alternative based on CMC slicing is to be interpreted as follows.
In some circumstances the CMC slicing is a privileged but changeable foliation. 
In others, it may instead be a completely absolute simultaneity. 
I.e. a stack of CMC leaves that cannot be viewed in any other manner, in direct parallel with the absolute simultaneity of Newtonian mechanics. 
The former possibility amounts to finding GR {\sl a second time} ($p/\sqrt{h} = constant$ and then $w = 1$ guarantees the refoliability) 
whilst the latter possibility gives a family of alternative theories.

Also note that since $\phi$ is trapped in the above equation, it is rendered a specifier equation rather than a constraint. 
I.e. a differential of the instant fixing equation (temporally relational version of an equation of a more commonly studied type called a lapse-fixing equation).

Various further interpretations of such theories/formulations are in \cite{ABFO, ABFKO, BO10, B11, Linking}.  
Approaches related to pursuing the option of interpreting the CMC condition as a further constraint include the `linking theory' shape dynamics program \cite{Linking}. 
This uses an enlarged phase space and trades symmetries. 
Here one deals with the second-class constraints/gauge fixings by the technique \cite{HT} of enlarging one's phase space.  
The present paper considers a more minimalist use of the cleaner (almost-)Hamiltonian mathematics, 
whereas these other papers consider an extended phase space use of this cleaner mathematics.  
Finally, we pose the question of what is the Dirac brackets formulation for the conformal option, since that is another standard approach upon encountering second-class constraints.

\section{Matter partner equation}\label{Matter}

Let $\psi^A$ denote fundamental-field second-order minimally-coupled bosonic matter.
This covers in particular minimally-coupled scalars, electromagnetism, Yang--Mills theory and the corresponding scalar gauge theories.
Then $\d s_{\sg\sr\sa\sv-\psi} =$

\noindent $\sqrt{\d s^{\sg\sr\sa\sv-\st\sr\si\sa\sll}_{y, w}\mbox{}^2 + \sum_{\psi}y^{-1}_{\psi}\d s^2_{\psi}}$
(minimal coupling gives no metric--matter kinetic cross-terms so it decomposes in this blockwise manner).
Here $\d s_{\psi}^2 := M_{AB}\d \psi^A\d \psi^B$ 
for matter configuration space metric $M_{AB}$ that we take to be ultralocal in the metric and with no dependence on the matter fields themselves.
Also $W_{\sg\sr\sa\sv-\psi} := aR + b + \sum_{\psi}a_{\psi} U_{\psi}$ for $U_{\psi}$ minus the matter sector's potential, which can only depend on the spatial derivatives of the
spatial metric through the spatial Christoffel symbols.

For many purposes an equivalent starting point is 
\beq
{\cal H}_{x, y, y_{\psi}, a, a_{\psi}, b} := \left\{y\{h^{ik}h^{jl} - x h^{ij}h^{kl}/2\}\pi_{ij}\pi_{kl} + 
\sum\mbox{}_{\mbox{}_{\mbox{\scriptsize $\psi$}}}  y_{\psi}  M^{AB}  \Pi_{A}\Pi_{B}\right\}/\sqrt{h} - \sqrt{h}\left\{aR + b + 
\sum\mbox{}_{\mbox{}_{\mbox{\scriptsize $\psi$}}} a_{\psi} U_{\psi}\right\} = 0 \mbox{ } .
\eeq
Here, $\Pi_A$ are the matter momenta conjugate to the matter variables $\psi^{A}$.
For these models, changes in all the matter degrees of freedom do have the opportunity to contribute to the emergent time standard, 
$t^{\se\sm(\sJ\sB\sB)} = \int \d s_{\sg\sr\sa\sv-\psi}/\sqrt{2W_{\sg\sr\sa\sv-\psi}}$.

The Poisson brackets between the GR--matter constraints are then 
$$
\mbox{\bf \{} (  {\cal H}_{x, y, y_{\psi}, a, a_{\psi}, b} | \d J) \mbox{\bf ,} \, ({\cal H}_{x, y, y_{\psi}, a, a_{\psi}, b}|\d K)\mbox{\bf \}}
=  
\left(
ay \{ { {\cal M}_i^{\sg\sr\sa\sv-\psi} } + 2 {\{1 - x\} D_i p} \}
+ \sum_{\psi}
\left\{ 
\underline{ ay
\left\lfloor
\Pi^{\sfA}\frac{\delta\pounds_{\d{\underline{\sF}}}\psi_{\sfA}}{\delta\d{\mF}^i}
\right\rfloor } 
\right.
\right.
$$
\beq
\mbox{ } \mbox{ } \mbox{ } \mbox{ } \mbox{ } \mbox{ } \mbox{ } \mbox{ } \mbox{ } 
\mbox{ } \mbox{ } 
\left.
\left.
       \left.
- 2 a_{\psi}y_{\psi} \underline{M^{{AB}}\Pi_{{A}}  \frac{\pa \mbox{\sffamily U\normalfont}_{\psi}}{\pa \, \pa_i\psi^{{B}}   }    }
      \right\}
	   - 2\underline{y        
			\left\{
p_{jk} - \frac{x}{2}p h_{jk}
            \right\}
h_{il}
\sum_{\psi}a_{\psi}            
			\left\{
            \frac{\pa \mbox{\sffamily U\normalfont}_{\psi}}{\pa {\Gamma^c}\mbox{}_{jl}}h^{ck}
          - \frac{1}{2}\frac{\pa \mbox{\sffamily U\normalfont}_{\psi}}{\pa{\Gamma^c}\mbox{}_{jk}}h^{lc}
            \right\}    }
\right| 
\d J \overleftrightarrow{\pa}^i  \d K 
\right)
\mbox{ } .
\eeq
Here ${\cal M}_i^{\sg\sr\sa\sv-\psi}$ is the gravity--matter version of the momentum constraint.  
Also the `floor bracket' denotes the extent to which the variational derivative inside acts. 
The above-listed matter fields all have no Christoffel symbol terms in their potentials, by which the last underlined grouping drops out.

The bracket of ${\cal H}_{x, y, y_{\psi}, a, a_{\psi}, b}$  with ${\cal M}_i^{\sg\sr\sa\sv-\psi}$ is just the obvious 3-diffeomorphism Lie drag expression.

The bracket of $\scH_{x, y, y_{\psi}, a, a_{\psi}, b}$  with $\scM_i^{\sg\sr\sa\sv-\psi}$ is the obvious 3-diffeomorphism Lie drag expansion.

In the GR option, we get $ay = a_{\psi}y_{\psi}$ to make the matter wave equations between the first and second underlined terms. 
This corresponds to $c_{\sg\sr\sa\sv} = c_{\psi}$ for each $\psi$ and thus $c_{\psi} = c_{\psi^{\prime}}$ for any two $\psi$, $\psi^{\prime}$.  
Thus minimally-coupled matter fields have to share propagation speed and null cone with gravity, and consequently have to share these things with each other.
This {\sl derives} rather than assumes the SR Principle, as a consistency condition \cite{FileR}.
In the Galilean option, the matter has matching infinite propagation speed.
In the Carrollian option the matter has matching zero propagation speed. 
This was partly anticipated by Henneaux \cite{Henneaux79} and its matter sector is ultralocal in the sense of Klauder \cite{Klauder}.
Finally, one can also have (some of) the matter be oppositely degenerate to the gravitational sector.  


\noindent This Sec is to be contrasted with Teitelboim's \cite{T80} treatment of minimally coupled matter in the Hojman--Kucha\v{r}--Teitelboim approach.
There the constraints additively split into gravitational and matter parts, and each part obeys the same algebraic brackets closure separately. 
Furthermore, both Teitelboim and the RWR program \cite{RWR, AB} get the Gauss constraint of electromagnetism, Yang--Mills theory and related theories as an integrability of $\scH$. 


\section{Conclusion}

\subsection{This paper's results}

Special Relativity (SR) usually precedes General Relativity (GR). 
This is so in Einstein's historical route to these two theories. 
Also Dirac's position was that it is easy to write down SR Lagrangians, and this is usually {\sl demanded} of the Lagrangians that one considers (p4 of \cite{Dirac}).

However, the `relativity without relativity' program proceeds oppositely, by showing that most other actions do not work. 
A variant of the Dirac procedure is used, and this enforces such a consistency.\footnote{N.B. 
`Reject as inconsistent' {\sl is} one of the 4 options in Dirac-like procedures \cite{Dirac}.  
This is clear from Dirac's example that $L = x$ gives equations of motion $0 = 1$. 
Moreover, SR is not by itself a guarantor of consistency. 
This is clear from the example of $L = L_0 + \Lambda_nS_n$ for $n + 1$ functionally-independent relativistic scalars $L_0$ and $S_n$, 
where $n >$ number of degrees of freedom of the system.}  
%
Whether a group $\fG$ is compatible with a given configuration space $\fQ$ is here to be discovered as one goes along with the dynamical investigation.\footnote{Nonetheless, 
it is also straightforward to encode symmetries into relational/RWR formulations (Sec \ref{DAE}). 
In particular, this is so for internal gauge symmetries, and reflects that Configurational Relationalism is a fairly standard alternative formulation of gauge theory.}

The assumption of mere consistency is, moreover, weaker than that of assuming (the standard) relativity.
This is clear through geometrodynamics-assumed RWR giving not one but four consistent options. 
I.e. GR, fixed-background Galileo-Riemann theory, Strong Gravity, and CMC-sliced formulations/theories. 
The first three locally have Lorentzian, Galilean and Carrollian relativities respectively.
It is interesting for these three to arise together from constrained dynamics. 
It is also interesting that they arise alongside the CMC condition that could be interpreted as a notion of absolute simultaneity hidden within GR or other geometrodynamical theories.

Moreover, if only metrodynamics is assumed, five options ensue. 
Four are as above, except that the Galileo--Riemann and Strong Gravity options now have 5 degrees of freedom. 
The fifth is $U$-geometrodynamics, in which only unit-determinant preserving spatial diffeomorphisms are physically meaningless.

One can take the above two paragraphs to constitute an answer to Wheeler's question about the GR form of the Hamiltonian constraint.
I.e. this arises as one of but a few consistent options upon assuming just the structure of space.
With minimally-coupled matter included, by each matter species having to share its null cone with gravity, they have to share null cone with each other. 
Thus gravity is now {\sl the explanation for} the SR principles occurring locally.
This is an approach in which dynamics is a more primary alternative to spacetime as regards the formulation of physical laws.  
Since this answer to Wheeler's question ascribes primality to space rather than to spacetime and yet leads to the recovery of spacetime, 
it also constitutes resolution of the classical Spacetime Reconstruction Problem in the sense of reconstruction from space.

Which of Leibniz--Mach--Barbour and Einstein's routes to GR is `more direct'?
We find that the claim that it is the former is somewhat misplaced due to setting up GR having more aspects to it than the arguments made usually cover. 
The four crucial aspects for this discussion are as follows. 
\noindent 1) The Equivalence Principle and the tie between gravitation and inertia as implemented by the mathematics of the connection, $\Gamma^{\mu}\mbox{}_{\nu\rho}$.
\noindent 2) The notion of curvature and its relation to matter content. 
\noindent 3) Specific Leibniz--Mach criteria for time and space.
\noindent 4) Whether spacetime or space are primary.

Then Einstein's historical route stays within a spacetime setting, changes status of frames from SR inertial frames to local inertial frames that are freely falling frames.
This makes direct use of the connection $\Gamma^{\mu}\mbox{}_{\nu\rho}$ to pass locally to freely falling frames. 
Then consideration of curvature tensors is natural, and a law is obtained relating the Einstein curvature tensor to the energy--momentum tensor of the matter content.  
This accounts for the local inertial frame's on physical grounds, and the spacetime geometry is to be solved for rather than assumed 
This approach does not directly address Machian criteria for time and space.

On the other hand, in the Leibniz--Mach--Barbour approach, time and space are conceived of separately, and Leibniz--Mach criteria are directly applied to each. 
The notion of space is broadened from that of the traditional absolute versus relational debate arena so as to include geometrodynamics. 
(This is without assuming the specific geometrodynamics that is obtained from the 3 + 1 split of GR.)  
This approach leads however to the recovery of the latter, and its equations imply the spacetime geometry formulation, which is then very fruitful as per usual.  
SR then arise from this as an idealization that holds well locally in many parts of the universe.  
Having obtained split 

\noindent 4-curvature, it is natural to ask about whether 4-connections also play a role in the theory.
One can build the latter out of natural space-first primary elements as per Appendix D.    
Thus one arrives indirectly at the identification of local inertial frames with freely falling frames. 
(Some -- but not all -- aspects of the Equivalence Principle are already present in the geometrodynamical Equivalence Principle \cite{HKT76}.)
Thus there is a trade-off of Equivalence Principle directness for Leibniz--Mach consideration of space and time directness.

\subsection{Caveats on (further) matter results}

Further matter uniqueness and Equivalence Principle `derivation' claims, purported to rest on relational postulates alone in the RWR paper, in fact relied on mathematical simplicities.  
This is via understanding from spacetime-presupposed canonical formulations that the \cite{RWR, AB} matter term ans\"{a}tze were not diverse enough to include 
more complicated realizations of the relational principles.  
Such can 
\noindent i) attain parametrization irrelevance by linearities other than the square root of a second-order kinetic term.   
\noindent ii) It can also include metric--matter kinetic cross-terms.
Then the following considerations apply.  

\noindent 1) The `further matter uniqueness claims' concerned deriving gauge theory fail. 
This was demonstrated \cite{Lan, Than, Phan, Lan2} via the inclusion of various other vector field theories as more complicated relational actions.
One such example is Proca Theory.  

\noindent 2) Furthermore, these include examples illustrating that SR is not the only finite-propagation-mode outcome, and of Equivalence Principle violation.
These become possible once one ceases to be simple in the sense of `minimally coupled'.  
Without this, the matter and gravitation are more interlinked and both can be needed to establish the propagation modes.
\cite{RWR} itself already established, at a Referee's request, that in fact Brans--Dicke theory is also relational.
\cite{Lan2} provided local-SR-cone-violating and Equivalence Principle violating vector--tensor theories \cite{Lan2} that nevertheless obey Configurational and Temporal Relationalism. 
These are a subset of Jacobson's Einstein--Aether theories \cite{AET}.
The Equivalence Principle derivation claim was in particular unravelled.  
This was via finding \cite{Phan} that \cite{RWR}'s ans\"{a}tze tacitly assumed what had been known to be a `geometrodynamical' form of Equivalence Principle since the days of \cite{HKT76}.  
This tacit assumption entered via precluding the presence of the metric--matter kinetic cross-terms that are kinematically appropriate for non-minimally coupled tensor matter field.  
(\cite{RWR}'s Brans--Dicke extension contained those appropriate for that particular non-minimally coupled scalar field.)  
Additionally, different values of the Brans--Dicke parameter correspond to $w \neq 1$ becoming consistent via involvement of metric--matter cross-terms. 
Thus Brans--Dicke theory and other more complicated scalar--tensor theories are available not only as strong gravity limit resolutions of RWR but also as finite propagation 
speed alternatives to the GR outcome of RWR. 

\noindent 3) Fermions require a linear kinetic term $\mT_{\sll\si\sn}$ being additively appended to the product of square roots (\cite{Van, Lan, FileR}). 
This exemplifies non-simplicity i).  

We finally provide a counter-caveat, in that Temporal and Configurational Relationalism can be coherently extended to a theory of Background Independence as per \cite{BI}.
Then some of the other premises of which exclude all Einstein--Aether theories and therefore the above family of counterexamples.
As such, there is at least some chance that a subset of the matter claims that RWR attempted to rest upon Temporal and Configurational 
Relationalism alone could rest on a more complete notion of Background Independence instead.

\subsection{Extensions of RWR}

\noindent A) We first note that the present paper's work already includes the de Sitter and anti de Sitter counterparts of the Poincar\'{e} group on larger scales.  
This is via identifying $b = -2\Lambda$ for $\Lambda$ the cosmological constant.
   
\noindent B) The present paper's work also immediately generalizes to the $n$-$d$ case.

\noindent C) On the other hand, using arbitrary signature for the lower-$d$ manifold (usually the role played by space) works for a number of the simpler results. 
However, this runs into difficulties with the theorems involving relativistically-appropriate analysis \cite{ACM}.  
This leads to the following position.  

\noindent i) The one-time multiple-space dimensions signature has particular physical significance.
It is supported by suitable p.d.e. analysis (GR Cauchy Problem), and has specific philosophical significance in the present paper's approach. 
(Spatial and Temporal Relationalism are distinct and each is grounded on the {\sl detailed properties} of space and time. 

\noindent
These include time being specifically 1-$d$ and space possessing a positive-definite Riemannian metric.)

\noindent ii) Then the choice of spatial dimension 3 is for now made due to its minimalistic accordance with observations.

\noindent However, the relational principles themselves can be made for any choice of spatial dimension.  
Thus they do not impede speculation that, or mathematical modelling of, scenarios with spatial dimension $>$ 3.  

\noindent D) Carrying out the present program's brand of relational thinking in arenas including each of supersymmetry and extended objects of dimension smaller than that 
of space itself remain open questions \cite{FileR}.

\noindent E) The present program's brand of relational thinking is also applicable to Ashtekar variables type formulations \cite{FileR}.  

\noindent Further, and probably the most immediate, lines of work for the issues raised and techniques used in the present paper are as follows. 

\noindent F) The RWR working and conformal theories in inhomogeneous perturbations about minisuperspace. 

\noindent G) The quantum counterpart of the present paper in terms of a suitable quantum commutator bracket algebraic structure. 
The most accessible such is probably the semiclassical version of 1), e.g. for a slightly inhomogeneous cosmology.
This can begin to address Sec \ref{2FQ}'s specifically quantum Spacetime Reconstruction Problem subfacets.

\noindent H) Spacetime and space reconstructions from discrete or discrete-type models (i.e. continuum limits of discrete theories such as causal dynamical triangulation)
are also a relevant part of some Spacetime Reconstruction Problem subfacets.  
Thus there may be interest in discrete(-type) analogues of the present paper and F) and G) above.  

\mbox{ } 

\noindent {\bf Acknowledgements}  E.A. thanks close people for support. 
Karel Kucha\v{r}, Chris Isham, Julian Barbour and Niall \'{o} Murchadha for some inspiring seminars and discussions over the years. 
Jeremy Butterfield, Marc Lachi$\grave{\me}$ze-Rey, Malcolm MacCallum, Juan Valiente--Kroon, Don Page and Reza Tavakol for help with E.A's career.
Harvey Brown and Simon Saunders for an invitation to talk on some of these matters at Oxford. 
The Perimeter Institute for hospitality during part of this work, which stay was furtherly made pleasant by further discussions with Cecilia Flori and Philipp Hoehn.  
F.M. acknowledges Julian Barbour for originally pushing him to think about Relativity without Relativity and Matteo Lostaglio 
for asking if one could combine the diffeomorphism and conformal constraints together.
Perimeter Institute is supported by the Government of Canada through Industry Canada and by the Province of Ontario through the Ministry of Economic Development and Innovation. 
F.M.'s research was also partly supported by grants from FQXi and the John Templeton Foundation.

\begin{appendices}

\section{Temporal Relationalism incorporated Principles of Dynamics}

See Fig \ref{TRiPoD} for a summary of which Principles of Dynamics structures and formulations require modifying to comply with Temporal Relationalism and which are already invariant.
The red leg is supplanted by the green leg whilst the blue leg is unaffected: these are the ``already temporally relational" parts of the standard PoD.
The new green leg is powered by FENoS (free end vnotion of space) value variation. 
This covers free end point value variation for particles and free end spatial hypersurface value variation for fields (and more cases in between if there are other-codimension objects).
(Due to yet more complicated auxiliaries, \cite{ABFKO} and shape dynamics need a distinct green leg from that given here.)
`d' stands for `differential', and `A' for `almost', meaning that auxiliary entities are treated differently but all physical entities are still treated the same way.
Cyclic differentials are the obvious parametrization-less equivalent to cyclic velocities.
In particular as regards this paper's workings, not only time but also velocities,    Lagrangians and                  Routhians require supplanting, 
                                                     respectively with differentials, Jacobi(--Synge) arc elements and d-Routhians. 
[These are indeed just `d-Routhian' rather than `d{\it A}-Routhian', since Routhians are allowed to contain velocities, so Routhians already include `almost-Routhians' as a subset.]
We covered the dA replacements for constrained systems' overall Hamiltonians and Dirac procedures in Sec \ref{DAH}.
On the other hand, configurations, momenta, actions, Poisson brackets, and incipient (as-yet unconstrained) Hamiltonians are stronger in being already well formulated 
enough to comply with Temporal Relationalism without any further alteration.
 
{            \begin{figure}[h!]
\centering
\includegraphics[width=0.97\textwidth]{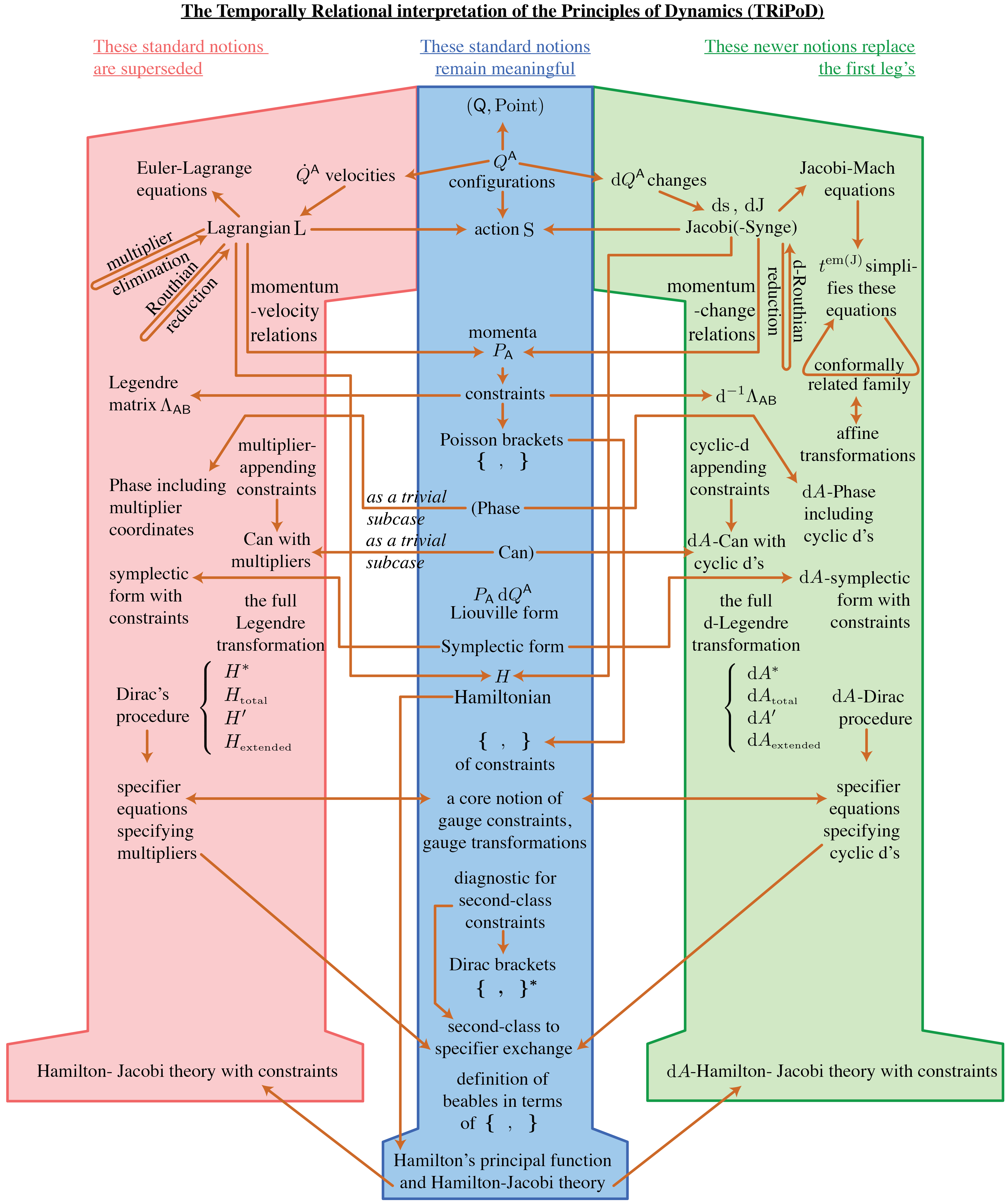} 
\caption[Text der im Bilderverzeichnis auftaucht]{        \footnotesize{The Temporal Relationalism implementation of the Principles of Dynamics (TRiPoD). 
Can denotes canonical transformations.} }
\label{TRiPoD}\end{figure}            }


\section{Logical completion of the thin sandwich in various formulations}

The BSW (\ref{S-BSW}), BFO--A (\ref{S-BFO-A}) and relational (\ref{S-rel}) actions all have the following `dual nationality'.
It is A) a (configuration-space-)geometrical presentations {\sl and} 
B) manifestations of spatial geometry as a carrier of information about time. (C.f. the BSW paper \cite{BSW}'s title.)   
This can be stated more precisely as `geometry and change of geometry is carrier of information about time'. 
Thus it is manifestly an implementation of Mach's Time Principle.

There are then various interesting layers of structure and concepts to comment upon.     

\noindent 1) The BSW action was set up as the start of the thin sandwich paradigm \cite{Wheeler63}.
Staying for now within BSW's formulation, the algorithm for this can be completed to the following.

\noindent 0)   Solve the Lagrangian-variables form of ${\cal M}_i$ with data ($h_{ij}, \dot{h}_{ij}$) for the shift $\beta^i$.
\noindent I)   One can then construct $\pounds_{\underline{\beta}}h_{ij}$, 
\noindent II)  then $\delta_{\vec{\beta}}h_{ij}$,
\noindent III) and then an emergent counterpart to $\alpha$, $N := \sqrt{\mT_{\sB\sS\sW}/4R}$.
\noindent IV)  Furthermore, an expression for an emergent time readily follows from construction III).
\noindent V) II) and III) permit one to construct the extrinsic curvature $K_{ij} = K_{ij}(x^k; h_{lm}, \beta^n, N]$. 
\noindent VI) III) permits one to construct an emergent version of the {\it tilt}, $\pa_b N$.
The suitability of this name \cite{Kuchar76} is most simply envisaged in the context of the standard formulation of SR.  
Here one has a family of boosted observers on a flat spatial surface tilted at a fixed angle to the undeformed flat spatial surface.   
\noindent VII) II), V) and VI) are the universal kinematics for split spacetime fields -- hypersurface derivative, extrinsic curvature `derivative coupling'\footnote{It 
is so named through featuring in metric-matter kinetic cross-terms in theories with non-minimially coupled matter fields \cite{Kuchar76}.}
and tilt -- one can construct a wide range of spacetime geometrical objects.
These include the spacetime connection (Appendix D), and, in the extension of the sandwich to tensor field matter, the spacetime form taken by that tensor field matter.

In contrast, the original sandwich prescription of Baierlein--Sharp--Wheeler involved just finding 0) and then [implicitly via I)] III) and then V).
IV) was first added to the procedure in Christodoulou's `chronos principle' work \cite{Chronos}.  
It was then separately discovered and exploited by Barbour \cite{B94I}. 
Object III) was geometrically identified as the hypersurface derivative by \K \cite{Kuchar76}, who also worked out the universal kinematics grouping of VII). 
The proposal that the sandwich algorithm should give as final output all of the universal kinematics rather than just the extrinsic curvature is original to the present paper.  
It represents a logical completion of the algorithm from the computational perspective.
Finally, the formula for extrinsic curvature used here is as per (\ref{lis}) except that the emergent $N$ has taken the place of ADM's presupposed $\alpha$.

\noindent 2) The thin sandwich algorithm can likewise be set up starting from the Routhian reduction of the {\it A}-action (\ref{S-A}) to form the BFO--A action.
This runs as above in terms of an auxiliary $\dot F^i$ in place of $\beta^i$ and $\dot{I}$ in place of $N$ as emergent quantity. 
The universal kinematics analysis indeed carries over to this formulation too, with emergent tilt now taking the form $\pa_a\dot{\tau}$.  

\noindent 3) In the Machian approach to GR assumed, one has a priori the BFO--A action without any need for a multiplier elimination or Routhian reduction.
The thin sandwich algorithm then mathematically runs as above, however it has a number of further lucid conceptual differences, so we write out the algorithm again in terms of these.

\noindent 0) The starting-point now amounts to solving the Jacobi variables ($Q^{\sfA}$, $\d Q^{\sfA}$)  form of ${\cal M}_i$ 
                             with Machian data ($h_{ij}, \d {h}_{ij}$) for the differential of the frame auxiliary $\d F^i$.   

\noindent Then one can construct the following.
\noindent I) $\pounds_{\d \underline{F}}h_{ij}$.  
\noindent II) The best-matching derivative $d_{\underline{F}} h_{ij}$ is the new conceptualization in place of the hypersurface derivative $\delta_{\vec{\beta}}$.
\noindent III) An emergent differential of the instant quantity 
\beq
\d I = {||\d_{\underline{F}}\bh||_{\sbM_{x,y}}}\left/{2\sqrt{ \sqrt{h} \{a R + b \} }}\right. \mbox{ } .
\label{gue}
\eeq
\noindent IV) An emergent time readily follows simply from integrating up III).
N.B. that this now constitutes abstracting a time-label for each instant, from a suitable notion of time. 
I.e. it is an implementation of Mach's Time Principle that furthermore allows for all change in principle for the geometrodynamical case but in practise keep only the locally relevant change. 
Thus it is a GLET is abstracted from STLRC implementation according to the conceptualization explained in Sec \ref{eph}.  
\noindent V) II) and III) permit one to construct the extrinsic curvature in its $K_{ij} = K_{ij}(x^k; h_{lm}, \d F^n, \d I]$ formulation (\ref{lis}).  
\noindent VI) III) permits one to construct an emergent version of a differential version of the tilt, $\pa_b \d I$.  
\noindent VII) II), V) and VI) are this new formulation's version of the three universal kinematics for split spacetime fields. 
Thereby one can once again construct a wide range of spacetime geometrical objects as per above.

\noindent 4) Suppose finally that one is looking to reconstruct spacetime and recover GR from amongst more general possibilities as expressed in the 3-space language.
The subsequent realization of the thin sandwich algorithm has all of the above steps and interpretations with the following exceptions. 

\noindent i) The interpretation as extrinsic curvature is not to be presupposed, so the interpretation of V) is that one is forming the `comparison of change' object 
$C_{ab} = \d_{\underline{F}} h_{ab}/ \,\d I$.
It is, moreover, in principle a `comparison with all change' and in practise a `comparison with STLRC'. 
Then the momentum formulation is entirely unaffected by the (\ref{lis}, \ref{gue}) distinction.
ADM and relational momenta coincide in the $x = 1 = w$, $y = 0$ case for which they all exist. 
Thus in this case comparing the `ADM-momentum to $K_{ab}$ relation' and the `relational momentum to $C_{ab}$ relation' permit the identification of $C_{ab}$ and $2K_{ab}$.  
We are henceforth entitled in this $x = 1 = w$, $y = 0$ case to use the shorthand
\beq
\d_{\underline{F}} h_{ab}/ \d I = 2 K_{ab} \mbox{ } .  
\eeq
\noindent ii) The above SR example explains that the name `tilt' is also inspired by spacetime geometry.  
This this should also be terminology that is reserved until after the recovery of spacetime.
Until that point this quantity is to be envisaged rather as the `spatial gradient of the change of instant'.  
Then after that point, the `proper time labels instants' duality converts this to the usual notion of tilt as recognizable from the above simple SR realization.
This is via the link of the observers' clocks' proper time exhibiting a constant gradient over space on the flat hypersurface tilted at a fixed angle.  
{\sl Thus in the relationalism-first rather than spacetime-first approach, Kucha\v{r}'s three types of universal kinematics are to be thought of, rather, as 
the best-matching derivative, the comparison of change with STLRC, and the spatial gradient of the change of instant.}

\section{Projection/Embedding Equations}

\noindent There are two ways of viewing what are mathematically the same equations but represent two distinct conceptualizations and differ in strength of prior assumptions. 
I.e. `projections computed' versus `embedding into spacetime deduced'.  
Presupposing the 4-$d$ manifold, one can {\sl evaluate} curvature tensor projections as F(curvature; $\alpha, \beta^a, K_{cd}]$: 
\beq
R^{(4)}_{abcd} = R_{abcd} - 2\epsilon K_{a[c}K_{d]b} \mbox{ (Gauss equation) } ,
\label{GE}
\eeq
\beq
R^{(4)}_{\perp bcd} = -2\epsilon D_{[c}K_{b]a} \mbox{ (Codazzi equation) } ,
\label{CE}
\eeq
\beq
R^{(4)}_{a\perp{c}\perp} = 
\mbox{\Huge \{} 
\stackrel{\mbox{$\{-\delta_{\stackrel{\rightarrow}{\beta}}  K_{ab} - \epsilon D_aD_b \alpha \}/\alpha \mbox{ } \mbox{ } + \mbox{ } \mbox{ } 
                                                             K_a\mbox{}^cK_{cb} \mbox{ } \mbox{ (Ricci equation in ADM decomposition) }$   }   } 
															 {\{-\delta_{\stackrel{\rightarrow}{\d F}}          K_{ab} - \epsilon D_aD_b \d \tau \}/\d \tau            
		                                                   + K_a\mbox{}^cK_{cb} \mbox{ } \mbox{  (Ricci equation in A decomposition)  }   } .
\label{RE}
\eeq
Here $\epsilon$ is the normalized version of $-a$ for ordinary ($\epsilon = -1$) and Euclidean ($\epsilon =  1$) GR. 
$\perp$ stands for `perpendicular', and is additionally a convenient coordinate adaptation for such equations \cite{Kuchar76}.   
Also useful later on are, setting $\epsilon = -1$ to cover the GR case, 
\beq
G^{(4)}_{ab} = \mbox{\Huge \{} 
\stackrel{    \mbox{$\{\delta_{\stackrel{\rightarrow}{\beta}}  K_{ab} - h_{ab} \delta_{\stackrel{\rightarrow}{\beta}}K - D_bD_a\alpha + h_{ab}D^2\alpha\}/\alpha 
          \mbox{ } \mbox{ } \mbox{ } - \mbox{ } \mbox{ } \mbox{ } \{ 2K_a\mbox{}^cK_{cb} - K K_{ab} +\{K_{ij}K^{ij} + K^2\} h_{ab}/2\}  + G_{ab}$}  } 
         {\{\delta_{\d \underline{F}}             K_{ab} - h_{ab} \delta_{\d \underline{F}}K     - D_bD_a\d T + h_{ab}D^2\d T\}/\d T   
          - \{ 2K_a\mbox{}^cK_{cb} - K K_{ab} +\{K_{ij} K^{ij} + K^2\} h_{ab}/2\}  + G_{ab}   } \mbox{ }                                              , 
\label{Gab}
\eeq
\beq
G^{(4)}_{a\perp} =  D_bK^b\mbox{}_a - D_aK \mbox{ } ,
\label{G0i}
\eeq
\beq
G^{(4)}_{\perp\perp} = \mbox{$\frac{1}{2}$} \{R + K^2 - K_{ij} K^{ij} \} \mbox{ } . 
\label{G00}
\eeq
\mbox{ } \mbox{ } But in the relational approach, without assuming the prior existence of the 4-$d$ manifold, we {\sl discover} 
\beq
0 = {\{\d_{\underline{F}}             K_{ab} - h_{ab} \d_{\underline{F}}K  \mbox{ }  \mbox{ }  \mbox{ }    -    \mbox{ }  \mbox{ }  \mbox{ } D_bD_a\d I + h_{ab}D^2\d I\}/\d I 
                - \{ 2K_a\mbox{}^cK_{cb} - K K_{ab} +\{K_{ij} K^{ij} + K^2\} h_{ab}/2\}  + G_{ab} \mbox{ }  }                                              \mbox{ } . 
\label{Gab2}
\eeq
which we can set to be a $G_{ab}^{(4)}$ identifying $F^i$ as $X^i$, $\d I$ as $\d \tau$ and recasting the best-matched derivative 
$\d_{\underline{F}}$ as a 

\noindent hypersurface derivative $\delta_{\d \stackrel{\rightarrow}{X}}$.
This comes alongside equations that can be identified with $G^{(4)}_{a\perp}$ and $G^{(4)}_{\perp\perp}$ without any further need of conversion:  
\beq
G^{(4)}_{a\perp} = -{\cal M}_a/2\sqrt{h} = 0 \mbox{ } ,
\label{G0iM}
\eeq
\beq
G^{(4)}_{\perp\perp} = -{\cal H}/2\sqrt{h} = 0 \mbox{ } .
\label{G00H}
\eeq
(\ref{Gab2}, \ref{G0iM}, \ref{G00H}) together form  
$
G^{(4)}_{\mu\nu} = 0:
$
the temporally relational form of the vacuum Einstein field equations.  
More simply at the level of the action, as fixed up by the consistency conditions imposed at the level of the constraints,
\beq
\int \int_{\Sigma} \d^3x  \sqrt{ \bar{R}} \, \d \ms \mbox{ } \mbox{ is equivalent to action with integrand } \mbox{ } R + K_{ij} K^{ij} - K^2 
\eeq
which, for $\Sigma$ compact without boundary, is equivalent to the Einstein--Hilbert action with $g_{\mu\nu}$ split of Fig 1.c).

Note that the Ricci requires a foliation rather than just the one surface. 
Thus the constraint/evolution equation distinction is based on an underlying kinematical distinction prior to having any Einstein field equations.  
This point is well laid out in e.g. \cite{Gour}, and is clearly structurally significant in recovering spacetime from space.  

\noindent Also note that including a cosmological constant $\Lambda = - b/2$ or minimally coupled matter each constitute a straightforward and well-known extensions.

\section{From relationalism to the spacetime connection and beyond}

Einstein's approach to GR makes direct use of the 4-connection to explain local inertial frames' realization by freely-falling frames.
This notion of frame replaces SR's (global) inertial frames, and is physically tied to the Equivalence Principle. 
With all matter species freely falling in the same way, the idea that all of these move on a common spacetime geometry becomes apparent.  
Then one is interested in which geometry this possesses and on how geometrical properties are inter-related with matter properties, 
which leads to the curvature to energy--momentum relation of the Einstein field equations.

As outlined in the Conclusion, in the relational approach, 
whilst space and time's Leibniz--Mach relational character is directly addressed, the above inertial-frame and Equivalence Principle considerations are {\sl not} directly addressed.
The natural sequence here is to 1) ask about the spatial 3-geometry of the world. 
2) To discover contracted embedding equations that strongly suggest modelling of physics as a 4-$d$ spacetime manifold.  
Moreover, these equations contain ab initio spatial curvature, and, interpreted as embedding equations, imply spacetime curvature also. 
It is then a natural question whether the spacetime connection that the spacetime curvature tensors can also be taken to be built from has a useful and lucid physical interpretation.  
This arises in 3-space terms as (e.g. in $\perp$ formulation)
\beq
{\Gamma^{(4)}}^{c}\mbox{}_{ab} = {\Gamma}^{c}\mbox{}_{ab} 
\mbox{ } , \mbox{ }  \mbox{ } 
{\Gamma^{(4)}}^{\perp}\mbox{}_{ab} = K_{ab} 
\mbox{ } , \mbox{ }  \mbox{ } 
{\Gamma^{(4)}}^{a}\mbox{}_{b\perp} = -\epsilon {K_{b}}^a 
\mbox{ } , \mbox{ }  \mbox{ } 
{\Gamma^{(4)}}^{\perp}\mbox{}_{\perp b} = 0 
\mbox{ } , \mbox{ }  \mbox{ } 
{\Gamma^{(4)}}^{a}\mbox{}_{\perp\perp} = \epsilon\,\pa_b\{\d I\}/\d I 
\mbox{ } , \mbox{ }  \mbox{ } 
{\Gamma^{(4)}}^{\perp}\mbox{}_{\perp\perp} = 0 \mbox{ } ,
\eeq
which can be packaged together as $\Gamma^{(4) \, \mu}\mbox{}_{\nu\rho}$. 
One still then requires Einstein's insights to realize how this object accounts for local inertial frames and its ties to the Equivalence Principle. 
Now new insight is claimed in this regard. 
Rather, this 1) clarifies where there is contact in the relational approach these aspects, which were largely not mentioned in \cite{RWR}.
2) It clarifies that it is both natural and mathematically straightforward in the relational approach to finally arrive at the notion of spacetime 4-connection 
playing a possible part in one's physics, via it having been established that the spacetime 4-curvature plays a role.  

\end{appendices}




\end{document}